\documentclass[a4paper,12pt, twoside]{article}
\usepackage{amsmath,float,bm}
\usepackage{hyperref}
\usepackage{amssymb}

\usepackage{authblk,graphicx}
\usepackage[total={6.5in,8.75in},
top=1.2in, left=0.9in, includefoot]{geometry}

\begin{document}

\title{\textbf{Classical and Bohmian trajectories in integrable and non integrable systems}}

\author{G. Contopoulos and A.C. Tzemos \footnote{Corresponding Author: atzemos@academyofathens.gr} }
\affil{Research Center for Astronomy and 
Applied Mathematics of the Academy of 
Athens - Soranou Efessiou 4, GR-11527 Athens, Greece}

\maketitle

\abstract{In the present paper we study the classical and the quantum Hénon-Heiles systems. In particular we make a comparison between the classical and the quantum trajectories of the integrable and of the non integrable Hénon Heiles Hamiltonian. From a classical standpoint, we study theoretically and numerically the form of the invariant curves in the Poincaré surfaces of section for several values of the coupling parameter of the integrable case and compare them with those of the non integrable case. Then we study the corresponding Bohmian trajectories and we find that they are chaotic in both cases, but chaos emerges at different times.}

\section{Introduction}\label{sec1}

A fundamental problem in Quantum Mechanics (QM) regards the existence of chaotic behaviour in the 
quantum regime. A classical dynamical system exhibits chaos when it has a bounded phase space and its trajectories are highly sensitive on the initial conditions. However, the standard Quantum Mechanics (SQM) does not predict trajectories for quantum particles. Thus chaos cannot be defined in a similar way as in Classical Mechanics (CM) and most works following the standard  QM rely on studying the  quantum properties of the integrable and chaotic classical counterparts. According to the Bohigas-Giannoni-Schmit  conjecture \cite{bohigas1984characterization} the differences  of the eigenvalues of a system corresponding to an integrable classical system follow a Poisson distribution, while the differences of the eigenvalues corresponding to a nonintegrable classical system folllow a Wigner distribution \cite{haake1991quantum,gutzwiller2013chaos,wimberger2014nonlinear,robnik2016fundamental}. This has been established numerically in various cases. However, this rule may not apply to all cases. 

Quantum chaos has been studied in various physical systems: a) in classicaly  driven chaotic systems, a representative class  being the ionized hydrogen atoms inside microwave fields \cite{casati1987relevance,bayfield1989localization}, b) in kicked quantum chaotic systems \cite{prange1989experimental}, c) in Bose-Einstein condensates \cite{milburn1997quantum,gardiner2002quantum,mahmud2005quantum,li2021multiparticle},
d) in materials where relativistic quantum transport phenomena take place \cite{huang2018relativistic,ying2016enhancement}, e) in nuclear systems \cite{gomez2011many} etc.

Bohmian Quantum Mechanics (BQM) is an alternative interpretation of QM which predicts deterministic trajectories for the quantum particles, governed by a set of a first order in time differential equations, the Bohmian equations \cite{Bohm,BohmII,holland1995quantum,pladevall2012applied}
\begin{equation}\label{eqtr}
m_i\frac{dx_i}{dt}=\hbar{Im}\left(\frac{\nabla \Psi}{\Psi}\right).
\end{equation}
where $\Psi$ is the usual wavefunction that satisfies the Schr\"{o}dinger equation
\begin{equation}\label{se}
-\frac{\hbar^2}{2m_i}\nabla^2\Psi+V\Psi=i\hbar\frac{\partial \Psi}{\partial t}.
\end{equation}
BQM is a nonlocal, trajectory based, quantum theory which predicts the same experimental results as standard QM, provided that the Bohmian particles are distributed according to Born's rule $P=|\Psi|^2$. The nature and the observability of Bohmian trajectories has attracted much interest in the past, both from a theoretical and an experimental standpoint.

In fact, much work has been done in recent years on applications of the Standard and Bohmian Quantum Mechanics in experimental setups. For example, very important experimental work was done  by Kocsis et al \cite{kocsis2011observing}, Braverman and Simon \cite{braverman2013proposal} and Foo et al. \cite{foo2022relativistic}, after the theoretical work of Wiseman on the `weak measurements' \cite{wiseman2007grounding}.

%\textbf{Furthermore in recent years there was much interest on the foundations of Quantum Mechanics (see e.g. the papers by Daly et al \cite{daley2022experimentally} and  Hance et al. \cite{hance2022bell, hance2024comment} on the Bell inequeality, by  Bhattacharya \cite{bhattacharya2024perturbative} on the mathematical origin of Planck's constant, by  Gisin  \cite{gisin2021indeterminism} on the classical chaos and Bohmian Mechanics and by Pozas-Kerstjens et al \cite{pozas2023proofs} on the quantum non locality. A comparison of these works with our theoretical approach is among our interests for future work. }

 The nonlinear nature of the Bohmian equations (Eq.~\ref{eqtr}) enables the study of chaotic phenomena at the quantum level using the framework of classical dynamical systems theory, where chaos is characterized by high sensitivity to initial conditions in a bounded phase space  Thus Bohmian chaos is an open field of research in quantum foundations and has been studied by many authors \cite{iacomelli1996regular, frisk1997properties, makowski2000chaotic, cushing2000bohmian, makowski2001simplest,wisniacki2005motion,
wisniacki2007vortex, santos2024broglie}. For a review of our work see \cite{contopoulos2020chaos}, while for an interesting study on the relation between classical chaos and Bohmian Mechanics see \cite{gisin2021indeterminism}.

From the early days of BQM it was understood that chaos emerges when a Bohmian particle comes close to a nodal point of $\Psi$ (the point where $\Psi=0$).  However, as it was shown in \cite{efthymiopoulos2007nodal,efth2009,tzemos2018origin}, in the frame of reference of a moving nodal point $N$, there is a second stagnant point $X$ of the Bohmian flow which is unstable, the so called 'X-point'. $N$ and $X$ form a nodal point-X-point complex (NPXPC). Whenever a Bohmian trajectory comes close to a NPXPC it gets scattered by the point $X$ and  the local Lyapunov exponent (the so called `stretching number') experiences a positive shift\cite{efthymiopoulos2007nodal,efth2009}. The cumulative result of many such close encounters between the trajectory and the NPXPCs is the saturation of the Lyapunov exponent at a positive value, indicating chaos.  The NPXPC mechanism describes all the major spikes in the time series of  the Lyapunov exponent. However, later, we found that the stagnant points of the Bohmian flow in the inertial frame of reference, the 'Y-points', also produce  chaos but their contribution is, in general, weak \cite{tzemos2023unstable}. 
The combined study of the X-points and the Y-points accounts for the profile of the Lyapunov exponent of a typical Bohmian trajectory. Trajectories that stay away from the $X$ and $Y$ points are ordered (thus their Lyapunov exponent is zero).

Up to now most works in Bohmian chaos have focused on systems of non interacting quantum harmonic oscillators in 2 dimensions \cite{borondo2009dynamical}. It is remarkable that even in the absence of interaction terms in a classical integrable system, its Bohmian counterpart has both chaotic and ordered trajectories. In order to observe Bohmian chaos one needs, in general, non commensurable frequencies and an entangled state of the system \cite{parmenter1995deterministic,makowski2000chaotic, makowski2001simplest,horodecki2009quantum,zander2018revisiting}.

Thus while a typical Bohmian system has both ordered and chaotic trajectories as a classical Hamiltonian system, Bohmian chaos emerges in a different way than classical Hamiltonian chaos, which is produced by  the overlapping of the asymptotic curves of its unstable periodic trajectories \cite{rosenbluth1966destruction, contopoulos1967resonance}. 

In the present paper we compare in detail  the ordered and the chaotic trajectories of two representative clasical systems, one integrable and one non integrable, and of the corresponding quantum analogues. As representative systems we have chosed simple systems, two cases of interacting oscillators  with coordinates $x$ and $y$ and with equal angular frequencies along the $x$ and $y$ coordinates ($\omega_x=\omega_y=1$)  described  by the H\'{e}non-Heiles Hamiltonians
\begin{equation}
H=\frac{1}{2}(x^2+y^2+\dot{x}^2+\dot{y}^2)+\epsilon(xy^2\pm \frac{x^3}{3}),
\end{equation}
The corresponding velocities  are $\dot{x}$ and $\dot{y}$  while $\epsilon$ is the coupling strength parameter.
When the sign of $\frac{x^3}{3}$ is negative  we have the original Hénon-Heiles system which is nonintegrable. Conversely, the positive sign refers to a case where a second integral of motion exists and the system is integrable \cite{lakshmanan2012nonlinear}.

The Hénon-Heiles Hamiltonian has been one of the most well studied systems in the theory of classical chaos due to its simple form and its high complexity as regards the corresponding trajectories \cite{henon1964applicability,henon1983numerical,fordy1983hamiltonian,
fordy1991henon}. Thus it has been used as a standard system for studying the signatures of classical chaos in the physical properties of the corresponding quantum systems. Moreover several variations of the H\'{e}non-Heiles model have been used extensively in the field of chemical Physics in order to understand vibrational modes in molecular systems. \cite{nordholm1974quantum,waite1981mode,takatsuka2022quantum}.
 However, only a few works have been done from a Bohmian standpoint \cite{chattaraj1993quantum,sengupta1996quantum,chattaraj1998chaotic}.

The structure of the present paper is the following:
In Section 2.1 we discuss the classically integrable quantum Hénon Heiles system. We find the invariant curves of the trajectories of the classical system. This system of course has no chaos. But in the corresponding quantum system (Section 2.2) most trajectories are chaotic (except if $\epsilon=0$) although their chaotic behaviour is apparent only after a rather long time. Then we find the eigenvalues of the chaotic system, the differences of the successive eigenvalues and the statistics of these differences. These statistics can be approximated by a Poisson distribution (and not  a Wigner distribution).

In Section 3 we do the same work for the non integrable Hénon-Heiles system. In this case there is both order and chaos in the classical system (Section 3.1). Then in Section 3.2 we study the corresponding Bohmian system. We calculate its eigenvalues and their differences. The statistics of these differences are again approximated by a Poisson distribution and not by  a Wigner distribution. Finally, we find the remarkable result that the times needed for chaos to emerge are much shorter in the Bohmian case corresponding to the  integrable classical system than in the nontintegrable system.

In Section 4 we make a  comparison between the two cases and draw our conclusions.

\section{Integrable Hénon-Heiles Hamiltonian}

\subsection{Classical Case}
\begin{figure}
\centering
\includegraphics[scale=0.22]{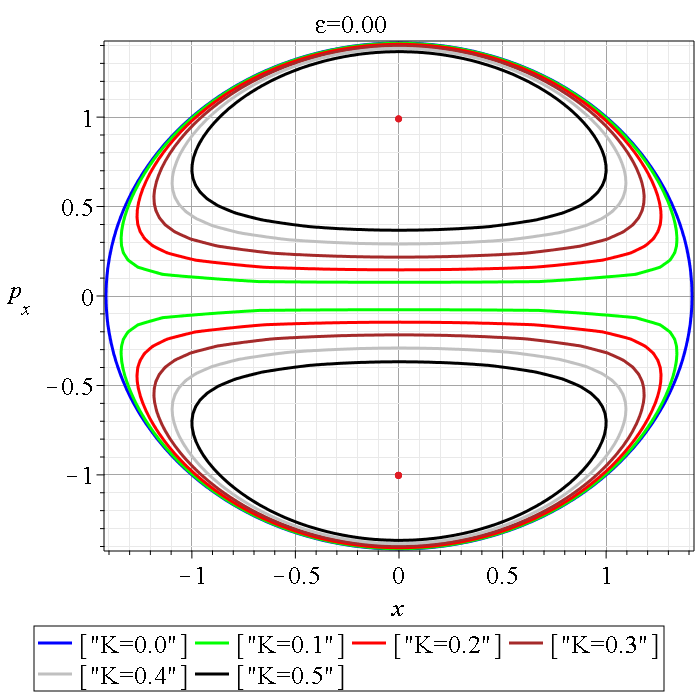}[a]
\includegraphics[scale=0.22]{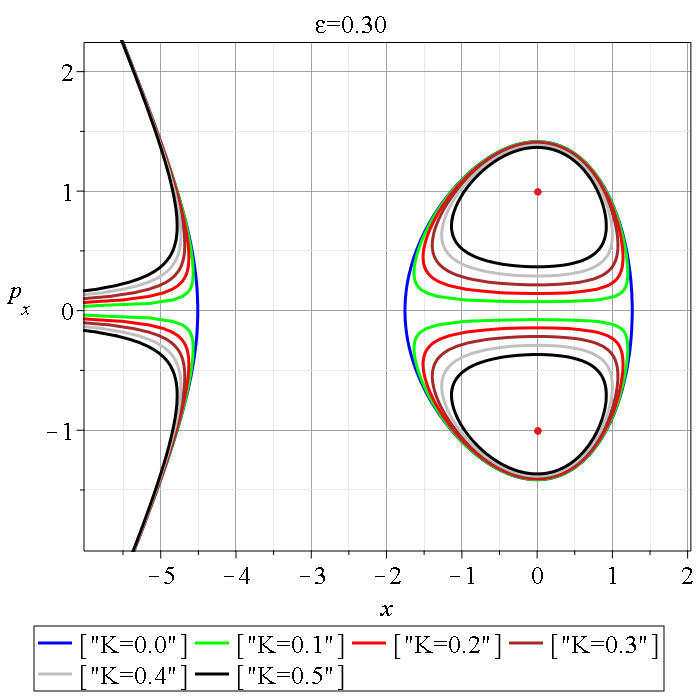}[b]
\includegraphics[scale=0.22]{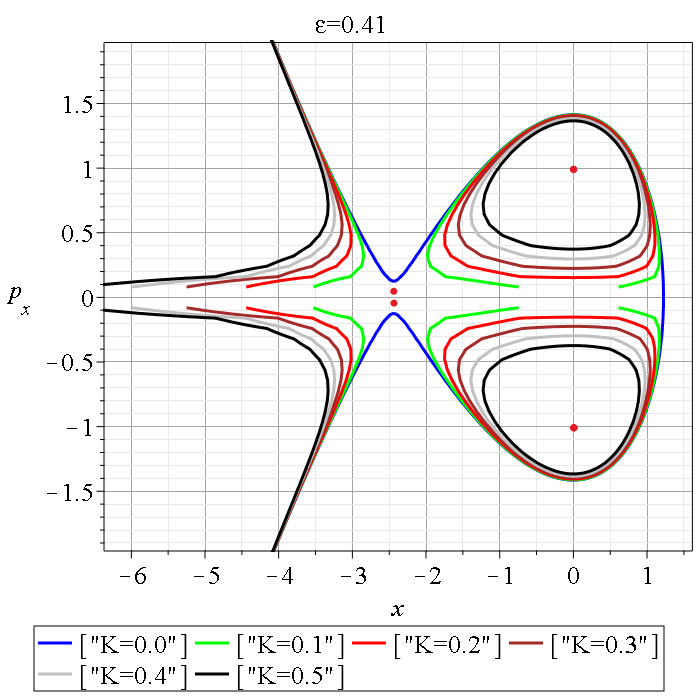}[c]
\includegraphics[scale=0.22]{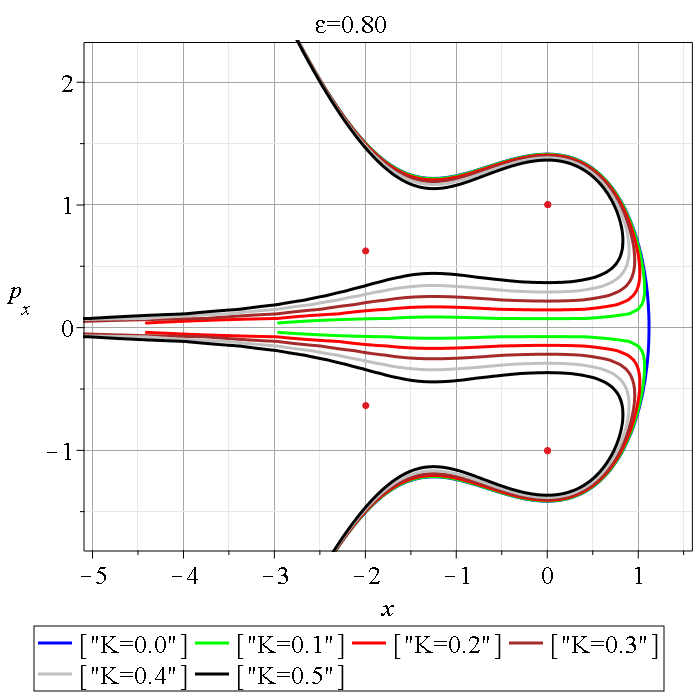}[d]
\includegraphics[scale=0.20]{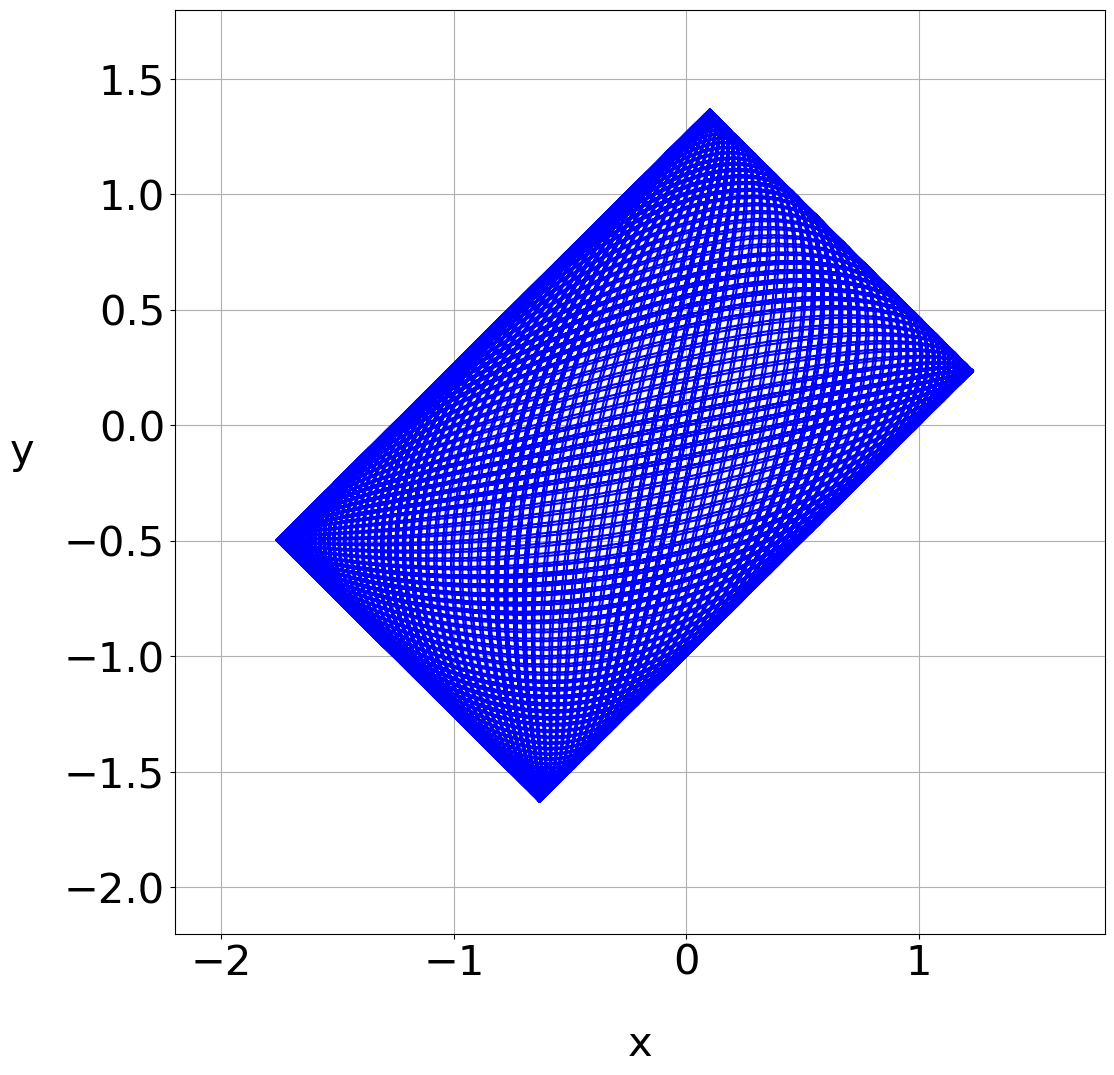}[e]
\includegraphics[scale=0.20]{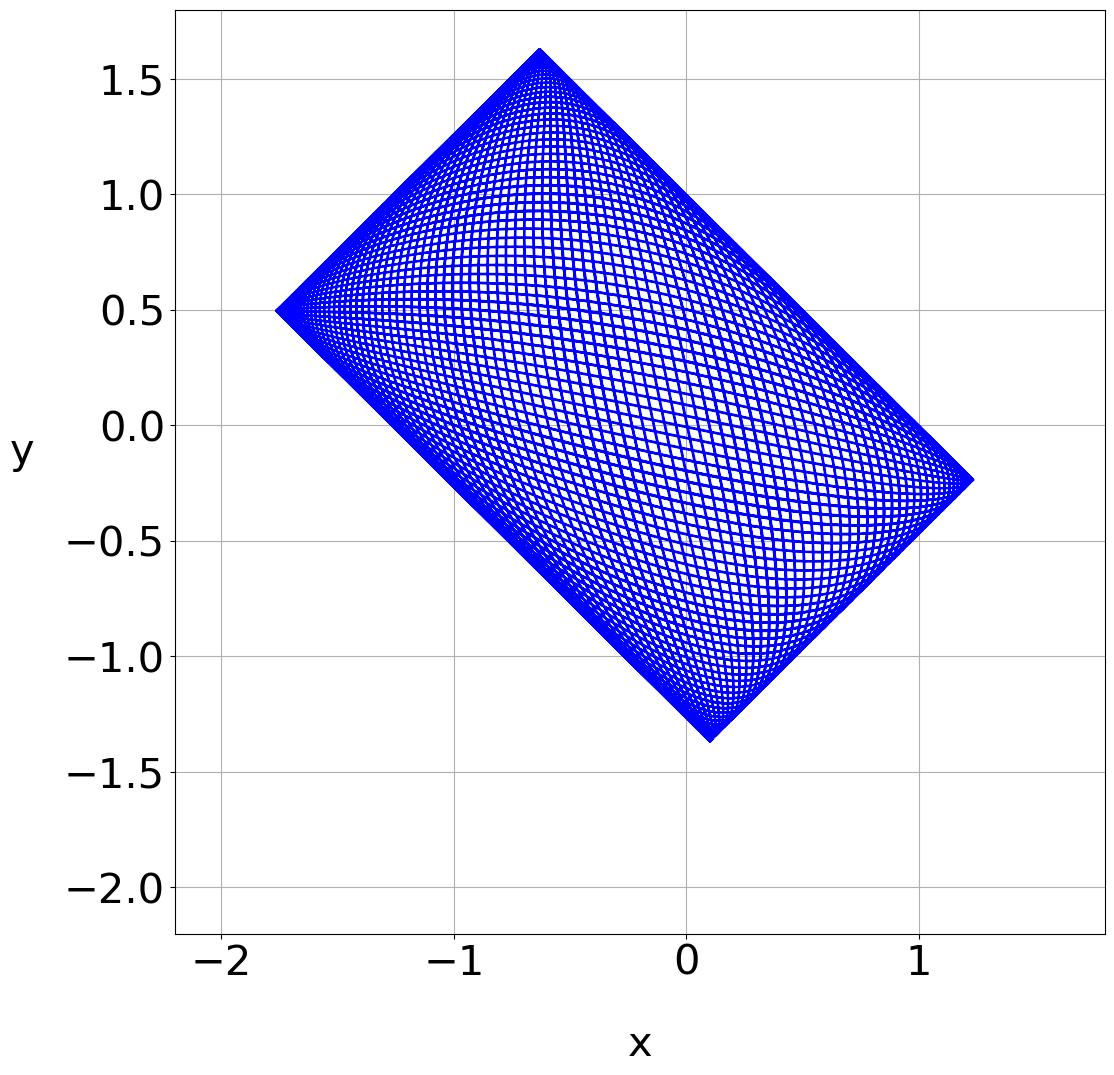}[f]\\
\includegraphics[scale=0.20]{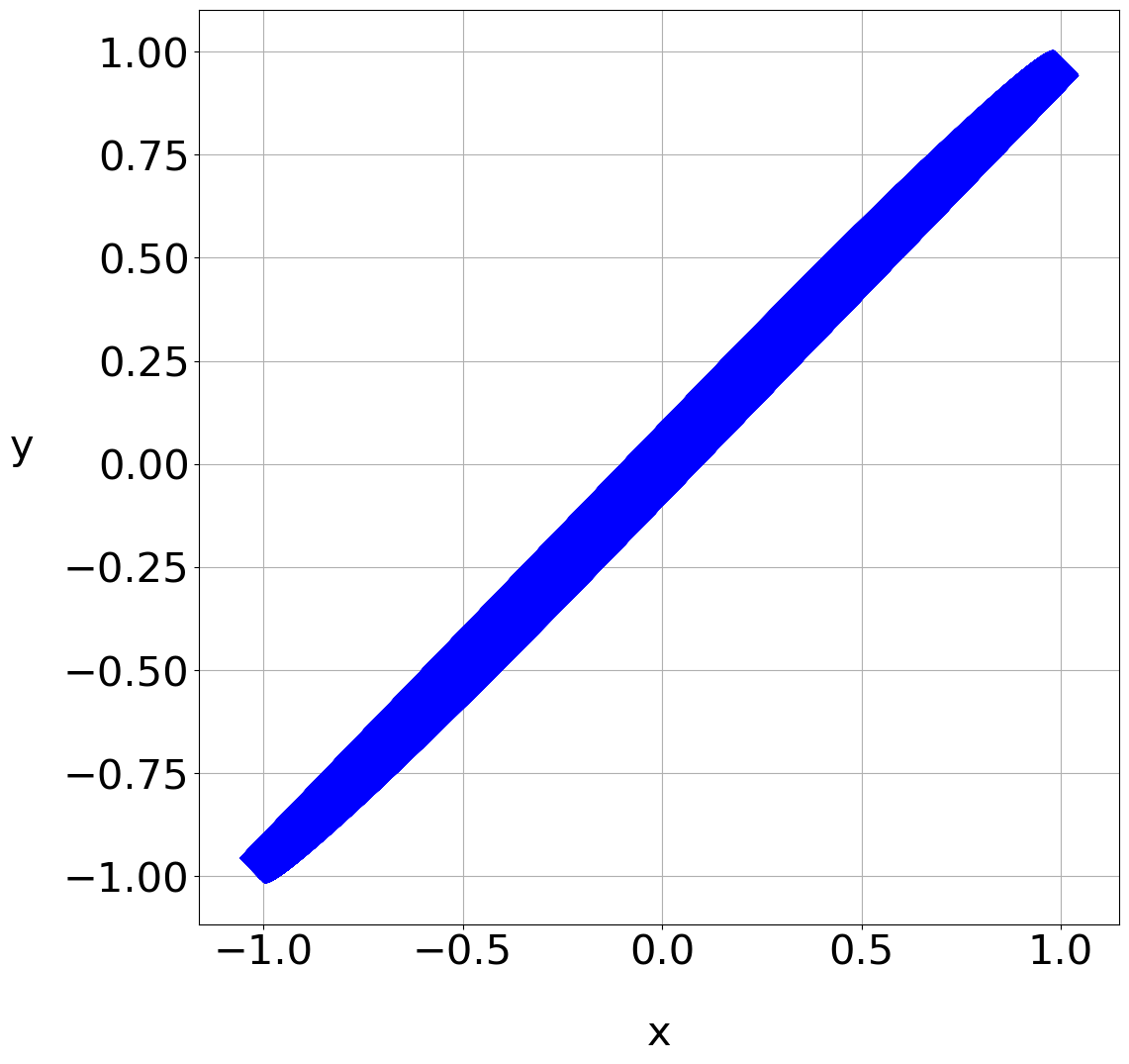}[g]
\includegraphics[scale=0.145]{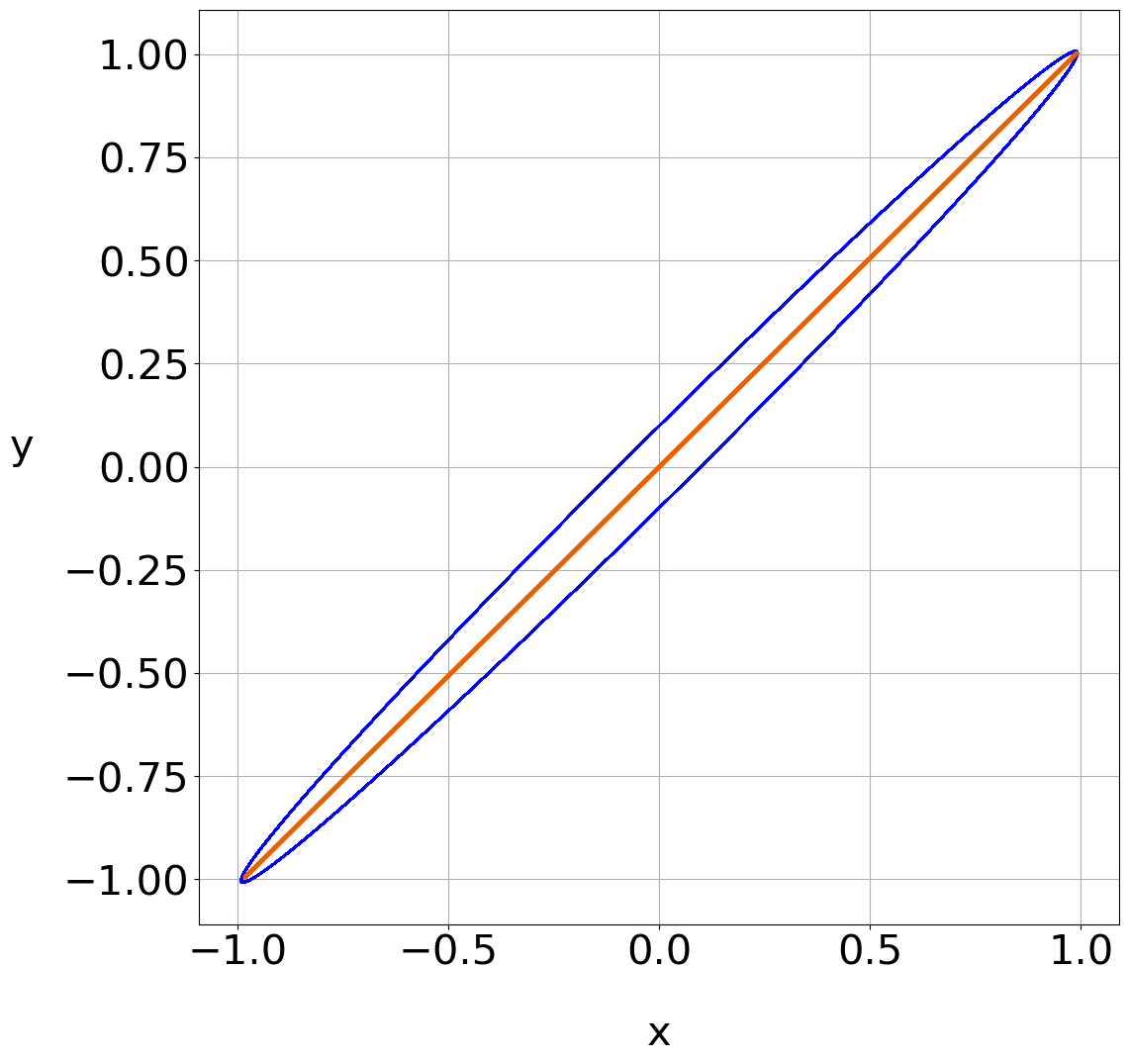}[h]
\caption{(a,b,c,d): Invariant curves for different values of the interaction term $\epsilon$ in the integrable case  for $E=1$. The different colors correspond to various values of the second integral $K$. In all cases $y=0$. a) $\epsilon=0$ b) $\epsilon=0.3$ c) $\epsilon=0.41$ and d) $\epsilon=0.8$. The red dots correspond to the periodic orbits, stable or unstable. Then we show the main types of non escaping classical orbits with: e) $x(0)=0.1, y(0)=0.2, \dot{x}(0)=-0.2768, \dot{y}(0)=-1.3678, \epsilon=0.3, K=0.4$, f) $x(0)=0.1, y(0)=0.2, \dot{x}(0)=0.3097, \dot{y}(0)=-1.3607, \epsilon=0.3, K=-0.4$, g) $x(0)=0.1, y(0)=0.2, \dot{x}(0)=\dot{y}(0)=0.9874, \epsilon=0.01, K=0.995$ and h) $ \epsilon=0, x(0)=0.1, y(0)=0.2, \dot{x}(0)=\dot{y}(0)=0.9874, K=0.9949$ (blue) and $\epsilon=0, x(0)=y(0)=0, \dot{x}(0)=\dot{y}(0)=1, K=1$ (red). }\label{fig1}
\end{figure}

The integrable classical  Hénon-Heiles Hamiltonian  system is described by the Hamiltonian:
\begin{equation}\label{H}
H=\frac{1}{2}(\dot{x}^2+\dot{y}^2+x^2+y^2)+\epsilon\left(xy^2+\frac{x^3}{3}\right)=E.
\end{equation}
It differs from the usual Hénon Heiles Hamiltonian only in the sign of the cubic term (+ instead of -). The corresponding second integral of motion is \cite{lakshmanan2012nonlinear}
\begin{equation}\label{Q}
Q=\dot{x}\dot{y}+xy+\epsilon\left(yx^2+\frac{y^3}{3}\right)=K.
\end{equation}
The invariant curves for $y=0$ are given by
\begin{equation}
\dot{x}\dot{y}=K,
\end{equation}
hence $\dot{y}=K/\dot{x}$ and Eq.~\ref{Q} gives 
\begin{equation}
f=\dot{x}^4+\left(x^2+2\frac{\epsilon x^3}{3}-2E\right)\dot{x}^2+K^2=0.
\end{equation}
Its solution is 
\begin{equation}
\dot{x}^2=\frac{1}{2}\left(2E-x^2-2\frac{\epsilon x^3}{3}\pm \sqrt{\left(2E-x^2-\frac{2\epsilon x^3}{3}\right)^2-4K^2})\right).
\end{equation}
In the numerical examples we set $E=1$ and for any given $\epsilon$ we have invariant curves for various values of $K$ (Figs.~\ref{fig1}a,b,c,d). If we fix a different value of $E$ then we find similar results.

The outermost invariant curve (for $E=1$ and $K=0$) is given by
\begin{equation}
\dot{x}^2=2-x^2-2\frac{\epsilon x^3}{3}.
\end{equation}
We have also the solution $\dot{x}^2=0$.

For $\epsilon=0$ the outermost invariant curve is a circle $\dot{x}^2+x^2=2$ and the axis $\dot{x}=0$. Inside this circle we have invariant curves above and below the axis $\dot{x}=0$. For $K>0$ the invariant curves are 
\begin{equation}
\dot{x}^2=\frac{1}{2}(2-x^2\pm\sqrt{(2-x^2)^2-4K^2}).
\end{equation}
For $x=0$ and $\epsilon=0$ we have the maxima and minima $\dot{x}^2$ with $\dot{x}^2=1\pm\sqrt{1-K^2}$. As $\epsilon$ increases from $\epsilon=0$ (Fig.~\ref{fig1}a) the circle is distorted and we have also a second set of invariant curves, on the left, that go to minus infinity (Fig.~\ref{fig1}b).  
If 
\begin{equation}
\frac{\partial f}{\partial x}=\frac{\partial f}{\partial \dot{x}}=0,
\end{equation}
then we have periodic trajectories.
For $E=1$ we find
\begin{eqnarray}
&\frac{\partial f}{\partial \dot{x}}=2\dot{x}\left[2\dot{x}^2+x^2+\frac{2\epsilon x^3}{3}-2\right]=0\\&
\frac{\partial f}{\partial x}=2\dot{x}^2[x+\epsilon x^2]=0.
\end{eqnarray}
One solution is 
\begin{equation}
x=0, \dot{x}^2=1
\end{equation}
This gives two stable periodic trajectories at the centres above and below the origin with $K=1$. The invariant curves for $K$ between $0$ and $1$ surround the unstable points $x=0,\dot{x}=\pm 1$.

There are also escapes along the y-axis. In Figs.~\ref{fig1}a,b such  orbits appear close to the points $(x=0, \dot{x}\pm=1)$ when $\epsilon>\frac{1}{2\sqrt{3}}\simeq 0.289$.

 Another solution is 
\begin{equation}
x=-\frac{1}{\epsilon}, \quad\quad \dot{x}^2=1-\frac{1}{6\epsilon^2}.
\end{equation}
This gives two unstable trajectories above and below the x-axis with $K=1-\frac{1}{6\epsilon^2}$ which exist when $\epsilon\geq \frac{1}{\sqrt{6}}$.

The value $\epsilon=\epsilon_{esc}=\frac{1}{\sqrt{6}}$ is the escape perturbation. The invariant curves with $K$ between $0$ and $1-\frac{1}{6\epsilon^2}$ extend to minus infinity close to the $x=0$ axis and above the upper periodic point or below the lower invariant point, surround from the right the stable periodic points. The invariant curves with $K$ between $1-\frac{1}{6\epsilon^2}$ and 1 form loops around the stable periodic points.

As $\epsilon$ increases from $\epsilon=0$ the second set of invariant curves approaches the main set (around (0,0)) and for $\epsilon=\frac{1}{\sqrt{6}}\simeq 0.408$ the outermost invariant curves reach each other at a singular point $\dot{x}=0, x=-2.449$.

For $\epsilon=0.5$ we have two unstable trajectories at $x=-2$ and $x=\pm \frac{1}{3}$ (Fig.~\ref{fig1}d). For $K>1$ there are invariant curves only at the left of $x=-\frac{1}{\epsilon}$. The corresponding trajectories  are, in general, Lissajous figures with axes parallel to the diagonals $x=\pm y$ (Fig.~\ref{fig1}e,f).

Theoretically the boundaries of these figures are given by eliminating $\dot{x}$ and $\dot{y}$ from the Eqs.~\eqref{H}-\eqref{Q} of the integrals of motion $H$ and $Q$ and the relation \cite{contopoulos1965resonance}
\begin{equation}
J=\frac{\partial Q}{\partial \dot{x}}\dot{y}-\frac{\partial Q}{\partial \dot{y}}\dot{x}=0.
\end{equation}
This relation gives
\begin{equation}
\dot{y}^2-\dot{x}^2=0\to \dot{x}=\pm\dot{y}.
\end{equation}
Introducing these values in Eqs. (2) and (3) we find for $E=1$
\begin{equation}
\dot{x}^2=1-\frac{1}{2}(x^2+y^2)-\epsilon\left(xy^2+\frac{x^3}{3}\right)=\pm\left [K-xy-\epsilon\left(x^2y+\frac{y^3}{3}\right)\right].
\end{equation}
This equation for $x,y$ can be written in the form
\begin{equation}
(x\mp y)^2+\frac{2\epsilon}{3}(x\mp y)^3=2\mp 2K.
\end{equation}
The solutions give two straight lines parallel to the diagonal $x=y$ at distances depending on $(1-K)$ and two straight lines parallel to the diagonal $x=-y$ depending on the value of $(1+K)$. Thus in general the Lissajous figures are elongated (Fig.~\ref{fig1}g). In the particular case $\epsilon=0$, all the trajectories are periodic forming ovals and in the limiting cases $1\mp K=0$ straight lines (Fig.~\ref{fig1}h).

\subsection{Quantum Case}

The Schr\"{o}dinger equation \ref{se} with $V=V_0+V_p=\frac{1}{2}(x^2+y^2)+\epsilon(xy^2\pm x^3/3)$ is not analytically solvable. However, we can follow the  standard procedure \cite{merzbacher1998quantum} of the direct diagonalisation of the perturbed Hamiltonian, exploiting the fact that the unperturbed system is solvable analytically and thus its eigenstates for a complete basis $\{|\Phi_l\rangle\}$ in the Hilbert space. The eigenstates are  $|\Phi_1\rangle=|0\rangle_x\otimes |0\rangle_y\equiv|00\rangle$, $|\Phi_2\rangle=|01\rangle$, $|\Phi_3\rangle=|10\rangle,\Phi_4=|02\rangle\dots$
where $|0\rangle, |1\rangle\dots$ correspond to the energy eigenstates of the 2 quantum harmonic oscillator in the $x$ or $y$ direction. Following this procedure the Hamiltonian operator in the basis $\Phi_l\rangle$ becomes a matrix whose arbitrary element is written as:
\begin{equation}
H_{i,j}=\int_{-\infty}^{\infty}\int_{-\infty}^{\infty}\Phi_i^{\dagger}(H_0+V)\Phi_jdxdy.
\end{equation}
The diagonalization of the Hamiltonian matrix of the perturbed system  gives the corresponding eigenvalues and eigenvectors. Then it is straightforward to calculate the spacings between the energy levels.

The eigenvectors of the perturbed system can then be written as linear combinations of the eigenstates of the unperturbed system,i.e.
\begin{eqnarray}
|\tilde{\Phi}_{r}\rangle=\sum_l c_{lr}|\Phi_l\rangle.
\end{eqnarray}
Thus an initial wavefunction $|\Psi_0\rangle=|\Psi(t=0)\rangle$ leads to 
\begin{equation}
|\Psi(t)\rangle=\sum_r\exp(-i\tilde{E}_rt/\hbar)\langle \tilde{\Phi}_r|\Psi_0\rangle|\tilde{\Phi}_r\rangle=\sum_{l,l',r}c_{l'r}^*c_{lr}\exp(-i/\hbar \tilde{E}_rt)\langle\Phi_{l'}|\Psi_0\rangle |\Phi_l\rangle.
\end{equation}
We start  with a wavefunction   which has in the position representation  the form
\begin{equation}\label{wave}
\Psi_0=a\Psi_{0,0}+b\Psi_{1,0}+c\Psi_{1,1}
\end{equation}
where $\Psi_{m,n}\equiv\Psi_m(x)\Psi_n(y)$ and $\Psi_m(x), \Psi_n(y)$ are the 1-D energy eigenstates of the oscillators \cite{merzbacher1998quantum} in $x$ and $y$ coordinates respectively, i.e.
\begin{equation}
\Psi_{m,n}(x,y,t)=\prod_{q=x}^y N_q\exp\left(-\frac{\omega_qq^2}{2\hbar}\right)\exp\left(-\frac{i}{\hbar}E_{s}t\right)H_s\left(\sqrt{\frac{M_q\omega_q}{\hbar}}q\right),\label{mn}
\end{equation}
and we set  $a=b=c=1/\sqrt{3}$ (thus $a^2+b^2+c^2=1$). Moreover $s=m,n$ are  (integers) for $x$ and $y$ while  $H_s$  stand for the Hermite polynomials of order $s$. $N_q$ is the normalization constant $N_q=\frac{(M_q\omega_q)^\frac{1}{4}}{\pi\hbar\sqrt{2^ss!}} $ and  the energy corresponding to the term  $\Psi_{m,n}$ is
\begin{equation}
E_{m,n}=E_m+E_n=\left(\frac{1}{2}+m\right)\hbar\omega_x+\left(\frac{1}{2}+n\right)\hbar\omega_y.
\end{equation}
We work with $\omega_x=\omega_y=M_x=M_y=\hbar=1$.

The wavefunction \eqref{wave} is the most well studied in the field of Bohmian Chaos \cite{parmenter1995deterministic}, since it has the simplest form that can exhibit chaos \cite{makowski2001simplest}.

\begin{figure}[H]
\centering
\includegraphics[scale=0.18]{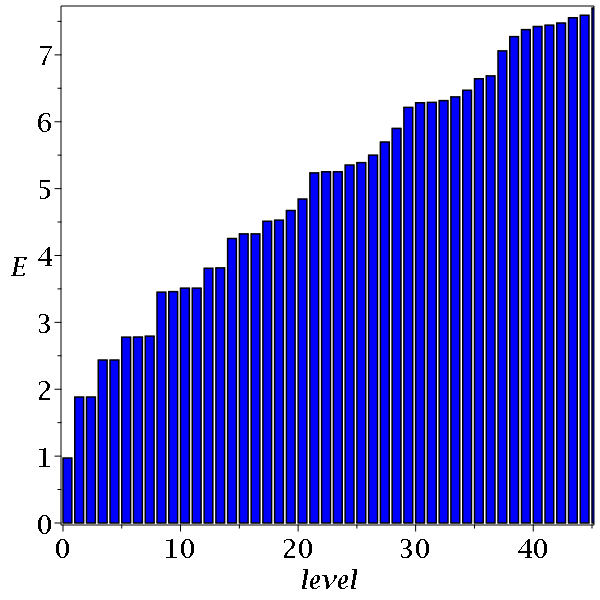}[a]
\includegraphics[scale=0.18]{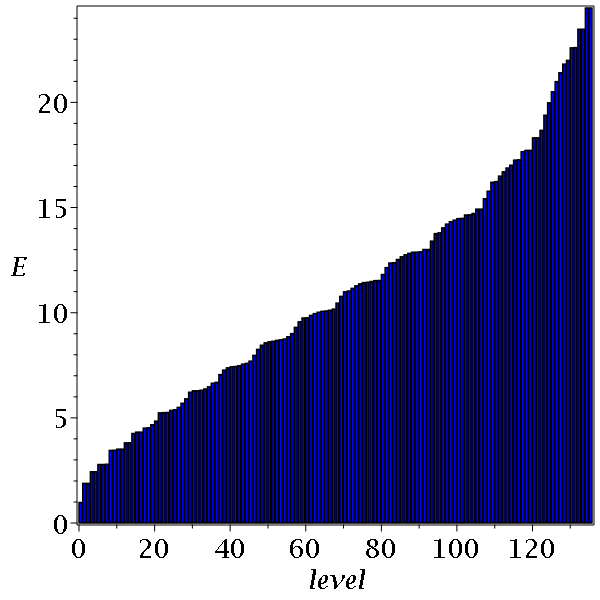}[b]
\includegraphics[scale=0.18]{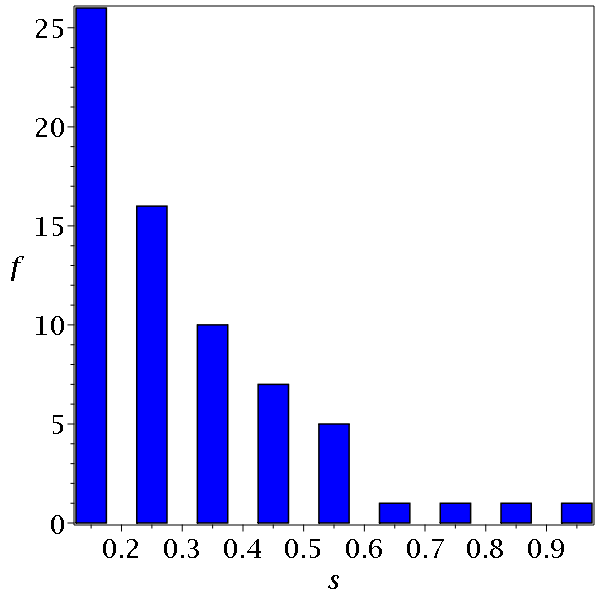}[c]
\caption{a) The distribution of the first 45 and b) of the first 136 energy  levels in the integrable case when $\epsilon=0.2$. c) The histogram of the (135) successive energy gaps of value $s$.}\label{fig3}
\end{figure}

We note that, due to the common frequencies, the unperturbed problem has a degeneracy, i.e. for a given value $n=n_x+n_y$ there are $n+1$ states with equal energies. This degeneracy is lifted by the term $\epsilon(xy^2+x^3/3)$. The number of states $|\Phi_l\rangle$ is truncated at a large level depending on $\epsilon$. The first 45 eigenvalues for $\epsilon=0.05$ are given in Fig.~\ref{fig3}a. The differences between the successive eigenvalues with the same sum $n=n_x+n_y$ are very small (the same level) while those between successive levels are large (Fig.~\ref{fig3}b). In this case ($\epsilon=0.05$) it is sufficient to take $N$ up to $N=45$ because the coefficients of the higher order terms are very small.
However, for larger values of the interaction parameter $\epsilon$ we need a much larger number of energy levels in our truncation.

For example, in the case $\epsilon=0.2$ it is sufficient to take a much larger number $N=136$ (Fig.~\ref{fig3}). If we calculate the differences between the successive energy levels of a given truncation then we can count how many times  we find a value $s$ in our ensemble. Such a histogram is given in Fig.~\ref{fig3}c. There we see that the distribution of the energy gaps is similar to  Poisson distribution (see also the conclusions section) of the form 
\begin{equation}
f(s)\varpropto \exp(-s),
\end{equation}
where $f$ is the proportion of the distances $s$ between adjacent energy levels.

Since the initial wavefucntion for  $\omega_x=\omega_y=1$ gives $E_{0,0}=1, E_{1,0}=2, E_{1,1}=3$, the total energy is $\langle E \rangle=a^2E_{00}+b^2E_{10}+c^2E_{11}=2$ and thus the average energy is $E_{av}=2$. For $\epsilon>0$ the average energy is larger than 2 and increases with $\epsilon$ (Fig.~\ref{fig12} in the last section).

The Bohmian trajectories in this case  are, in general, chaotic. This is due to the fact that the trajectories approach the moving nodal points and are deviated by their nearby X-points according to the nodal point X-point complex mechanism \cite{efthymiopoulos2007nodal,efth2009}.

This fact, however, is not obvious. In a previous paper \cite{tzemos2019bohmian} we considered the resonant case of a two qubit system of the unperturbed 2d harmonic oscillator, where we have in general an infinity of nodal points along a straight line. In that paper we found that when the frequencies $\omega_x, \omega_y$ of the unperturbed potential $V=\frac{1}{2}(\omega_x^2x^2+\omega_y^2y^2)$ are commensurable, then all the corresponding Bohmian trajectories are periodic. In particular when $\omega_x=\omega_y$ there are no nodal points at all (the analytical formulae for the positions of the nodal points have a denominator $\sin[(\omega_x-\omega_y)t]$ that becomes zero at $\omega_x=\omega_y$, i.e. the nodal points are at infinity).

In the present case we have the same resonance $\omega_x=\omega_y=1$. However in the quantum Hénon-Heiles problem the position of the nodal points has no specific form in the configuration space due to the anisotropy in the $x$ and $y$ coordinates. Namely, the nodal points enter the central region of the support of the wavefunction in a complex way and scatter the incoming Bohmian trajectories.

\begin{figure}[H]
\centering
\includegraphics[scale=0.22]{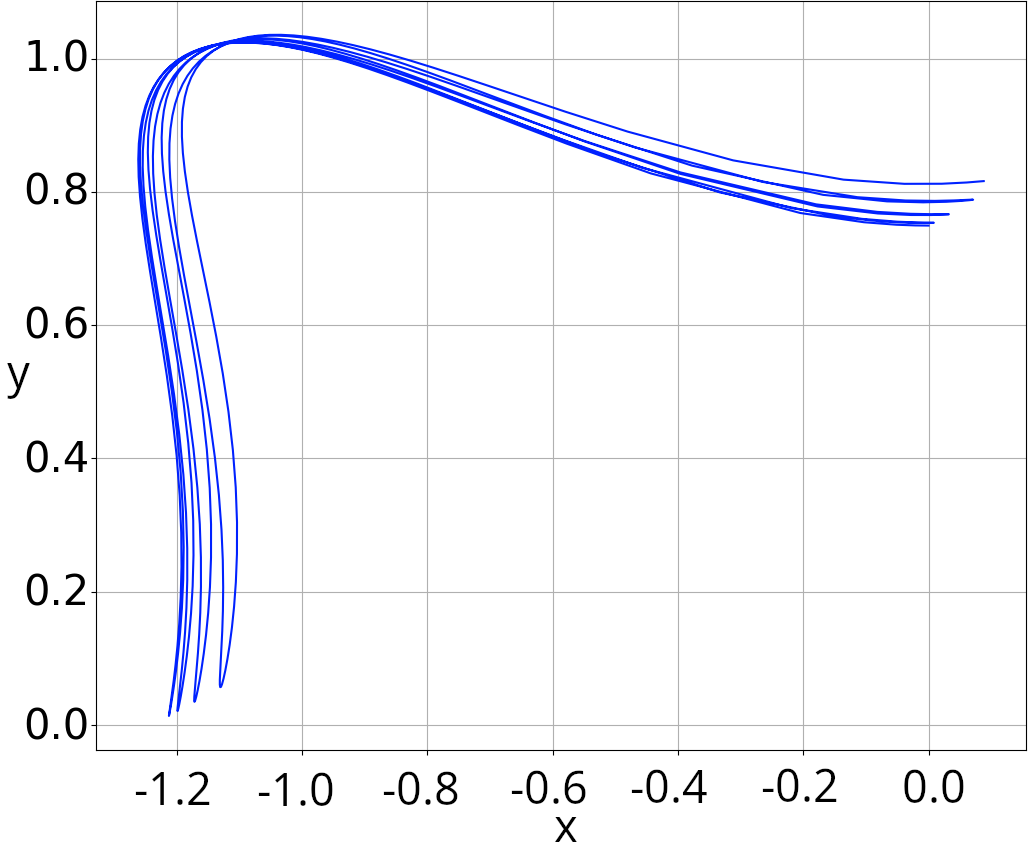}[a]
\includegraphics[scale=0.22]{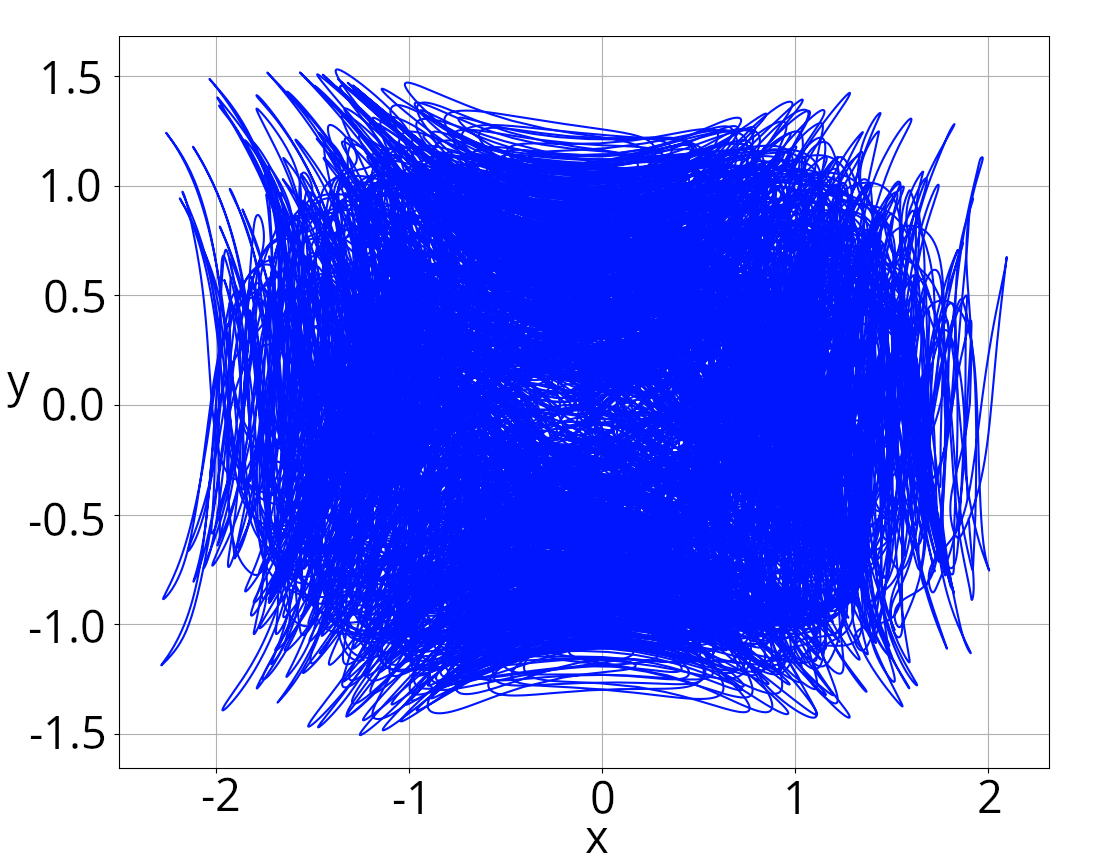}[b]
\includegraphics[scale=0.22]{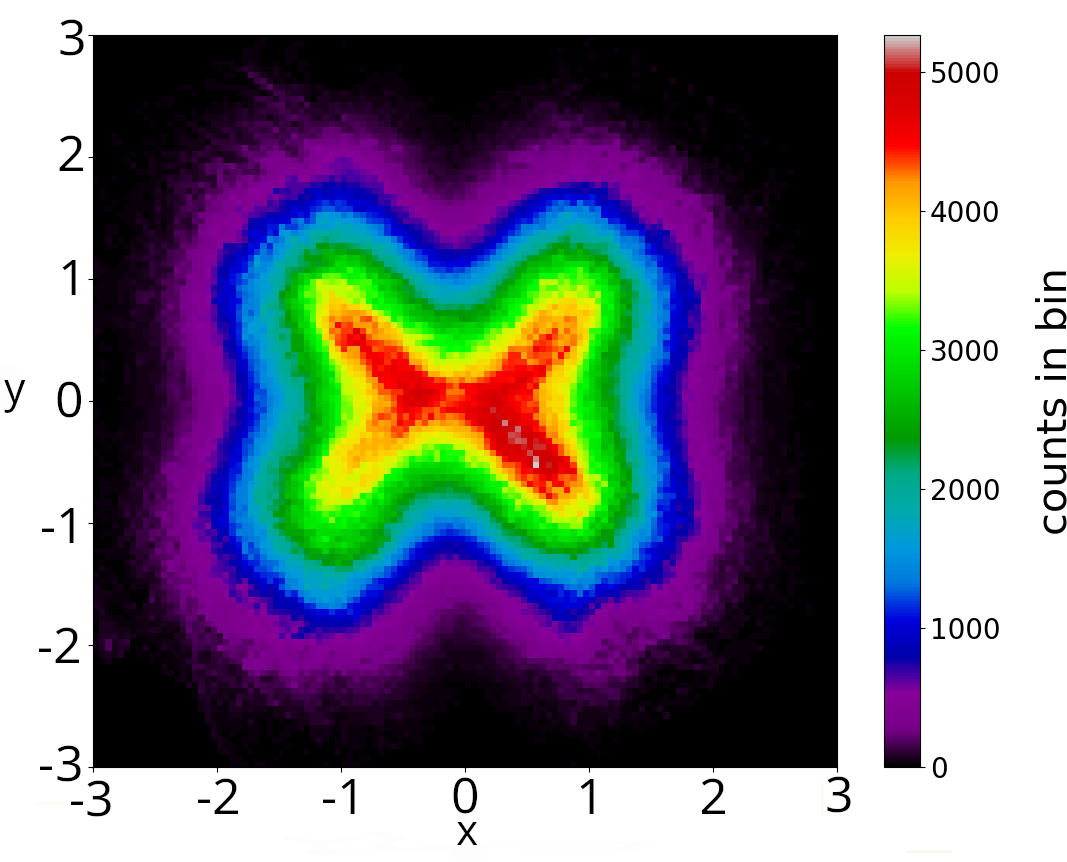}[c]
\includegraphics[scale=0.27]{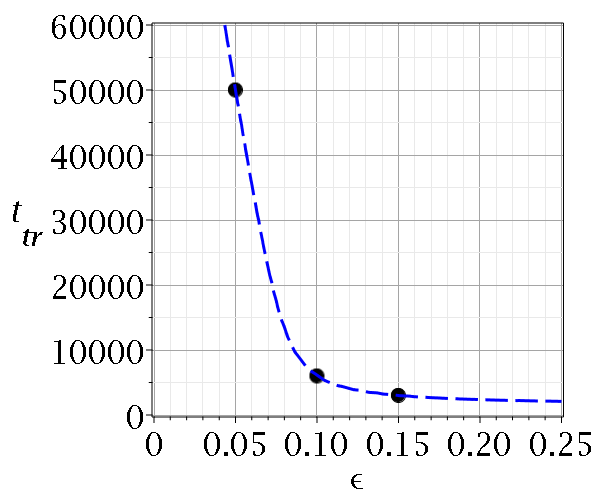}[d]
\caption{a) A Bohmian trajectory  ($x(0)=0, y(0)=0.75$) for $\epsilon=0.05$ up to a) $t=4\pi$ and b) $t=10^3\times 2\pi$. In c) we give the long limit distribution of the points of this trajectory up to $t=10^5\times 2\pi$. It is a chaotic-ergodic trajectory. Finally in d) we show the time needed for about $98\%$  of the trajectories, produced by  5000 Born-distributed initial conditions, to show their chaotic character for various values of $\epsilon$ (integrable case).}\label{fig4}
\end{figure}

In (Fig.~\ref{fig4}a) we calculate a Bohmian trajectory up to $t=4\pi$ for $\epsilon=0.05$. If this trajectory would be periodic it should return to its initial point after every $\Delta t=2\pi$. This does not happen in the present case. Therefore the trajectory is not periodic. The same trajectory was then integrated (for $\epsilon=0.05$) up to $t=2000\pi$.  This is  chaotic (Fig.~\ref{fig4}b), as it was verified by calculating the Lyapunov characteristic number. Its corresponding colorplot, which shows the number of times a trajectory passes through every bin of area $0.05\times 0.05$ in the configuration space up to $t=200000\pi$ is shown in Fig.~\ref{fig4}c. We see that the Bohmian trajectories are very different from the classical trajectories. They approach some nodal points from time to time and they become chaotic in the long run. In fact they require some time to exhibit their chaotic character that depends on the value of the perturbation parameter $\epsilon$.

The manifestation of chaoticity is made when the Bohmian trajectories cover practically all the available support of the wavefunction \cite{tzemos2020ergodicity,tzemos2021role}. Namely, as the Bohmian particles after the approach  of a nodal point they change their apparently ordered form  and extend to larger distances, showing their chaotic character. The transition time $t_{tr}$ when most trajectories  in an ensemble of 5000 Born-distributed initial conditions are shown to be chaotic (about $98\%$) is given for some values of $\epsilon$ in Fig.~\ref{fig4}d.

If the perturbation $\epsilon$ is large the time of manifestation of chaoticity is short, but as $\epsilon$ decreases this time is larger and becomes very large for small $\epsilon$, tending to infinity as $\epsilon\to 0$. This behaviour is similar to that  already observed for non commensurable frequencies \cite{tzemos2024comparison}.

\section{Nonintegrable Hénon-Heiles Hamiltonian}

\subsection{Classical Case}
The non integrable Hénon-Heiles Hamiltonian  is 
\begin{equation}
H=H_0+\epsilon H_1=\frac{1}{2}(\dot{x}^2+\dot{y}^2+x^2+y^2)+\epsilon(xy^2-\frac{x^3}{3})=E,
\end{equation}
which has both ordered and chaotic trajectories.

\begin{figure}[H]
\centering
\includegraphics[scale=0.2]{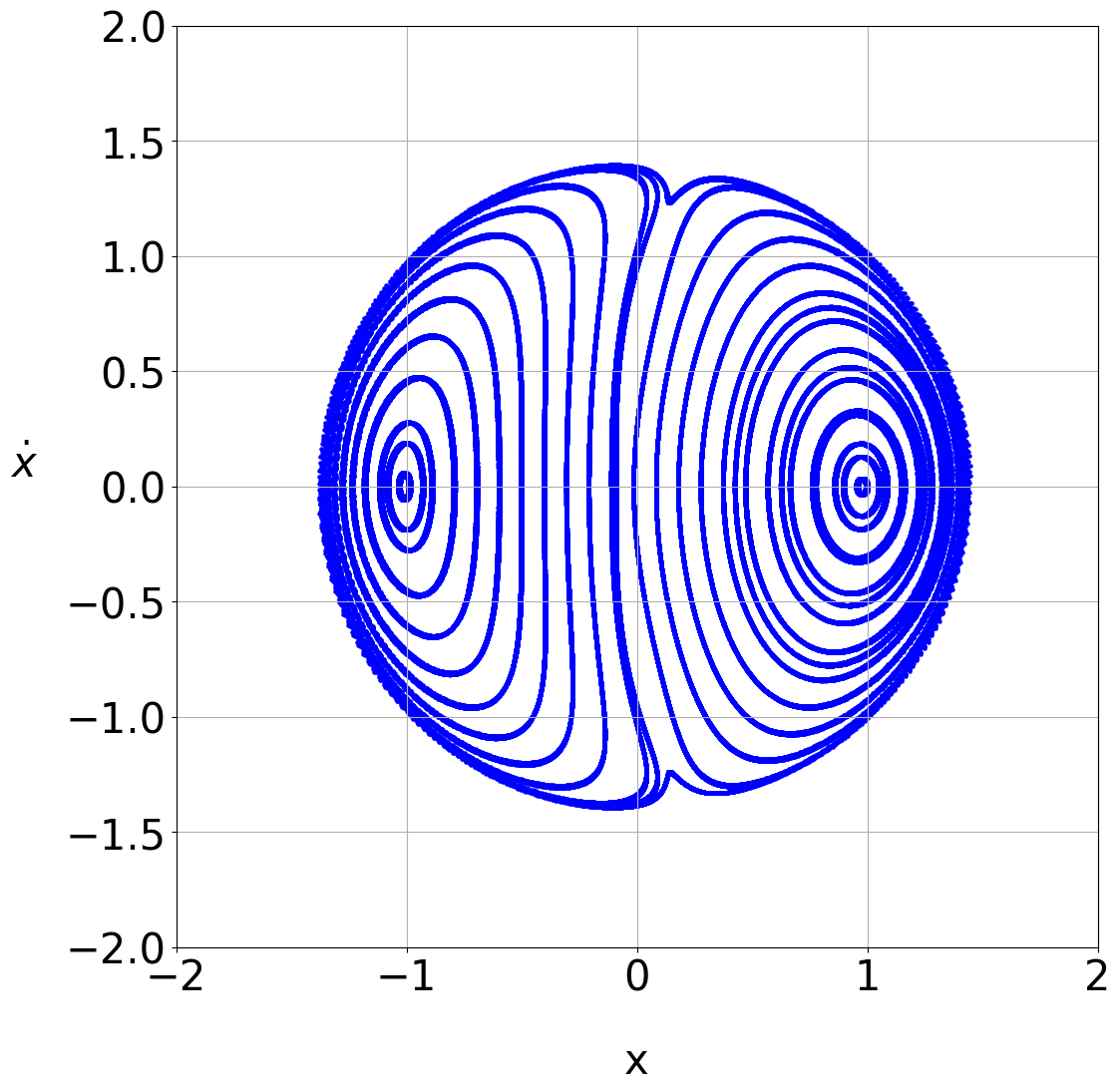}[a]
\includegraphics[scale=0.2]{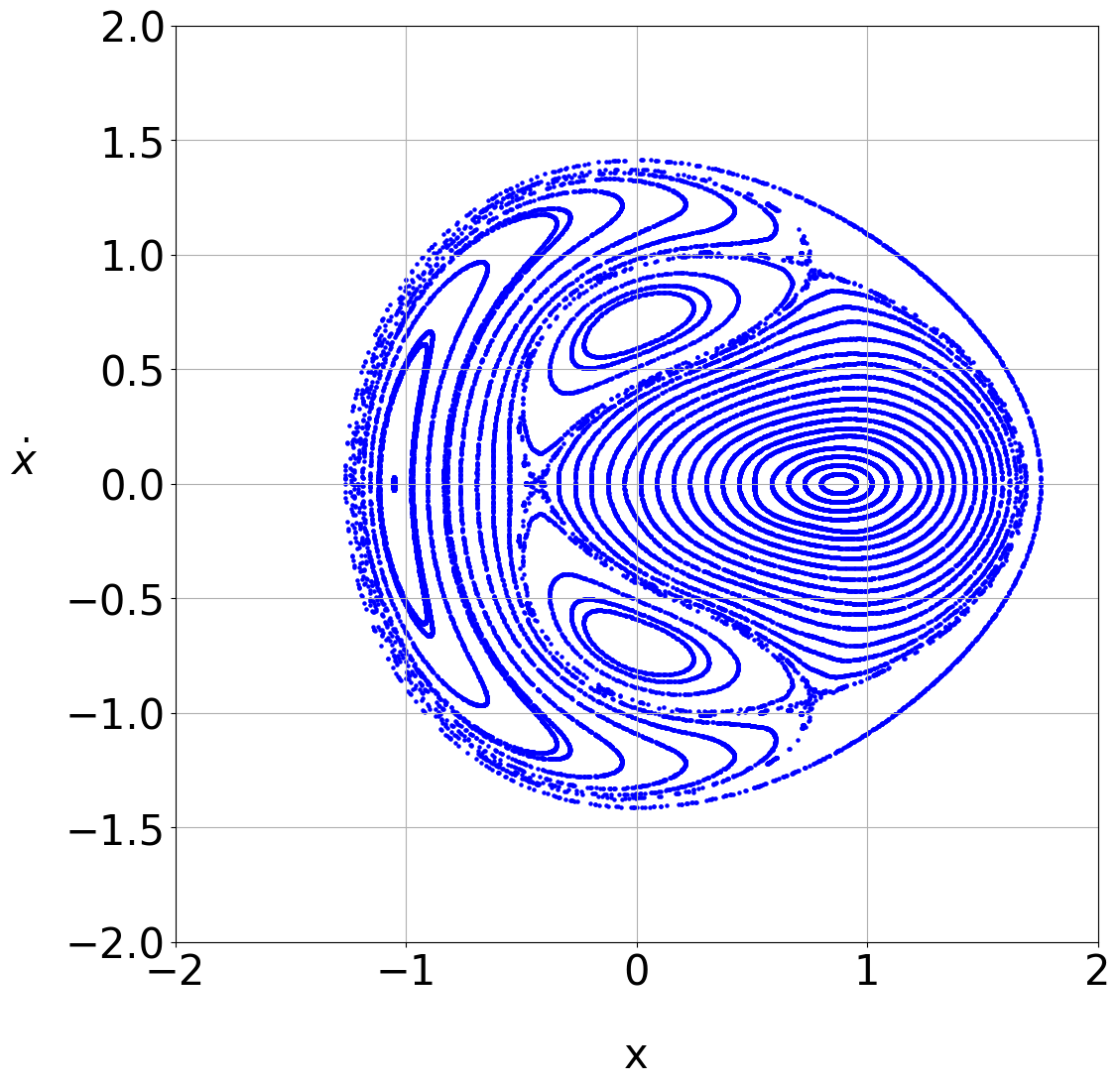}[b]\\
\includegraphics[scale=0.2]{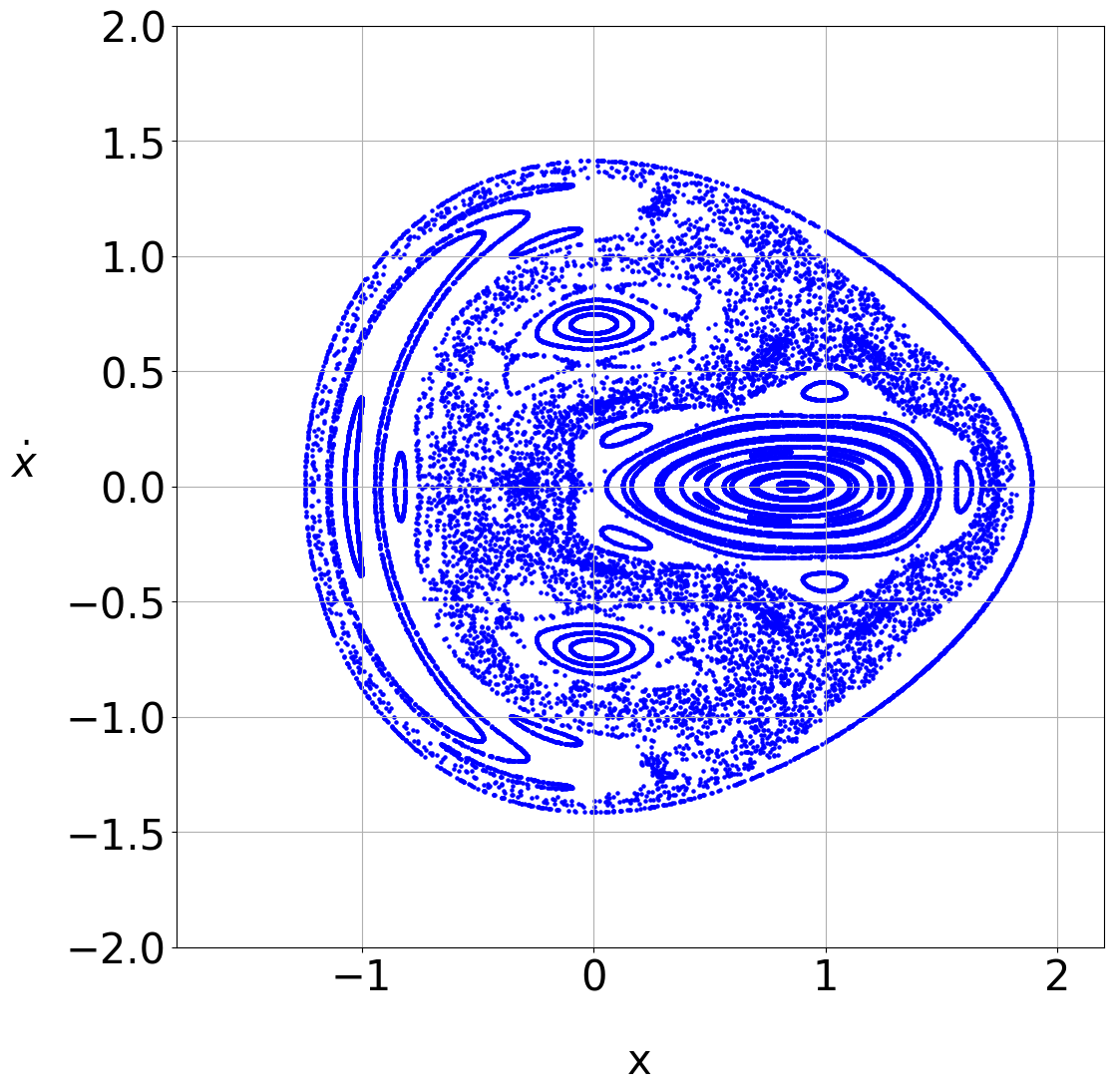}[c]
\includegraphics[scale=0.2]{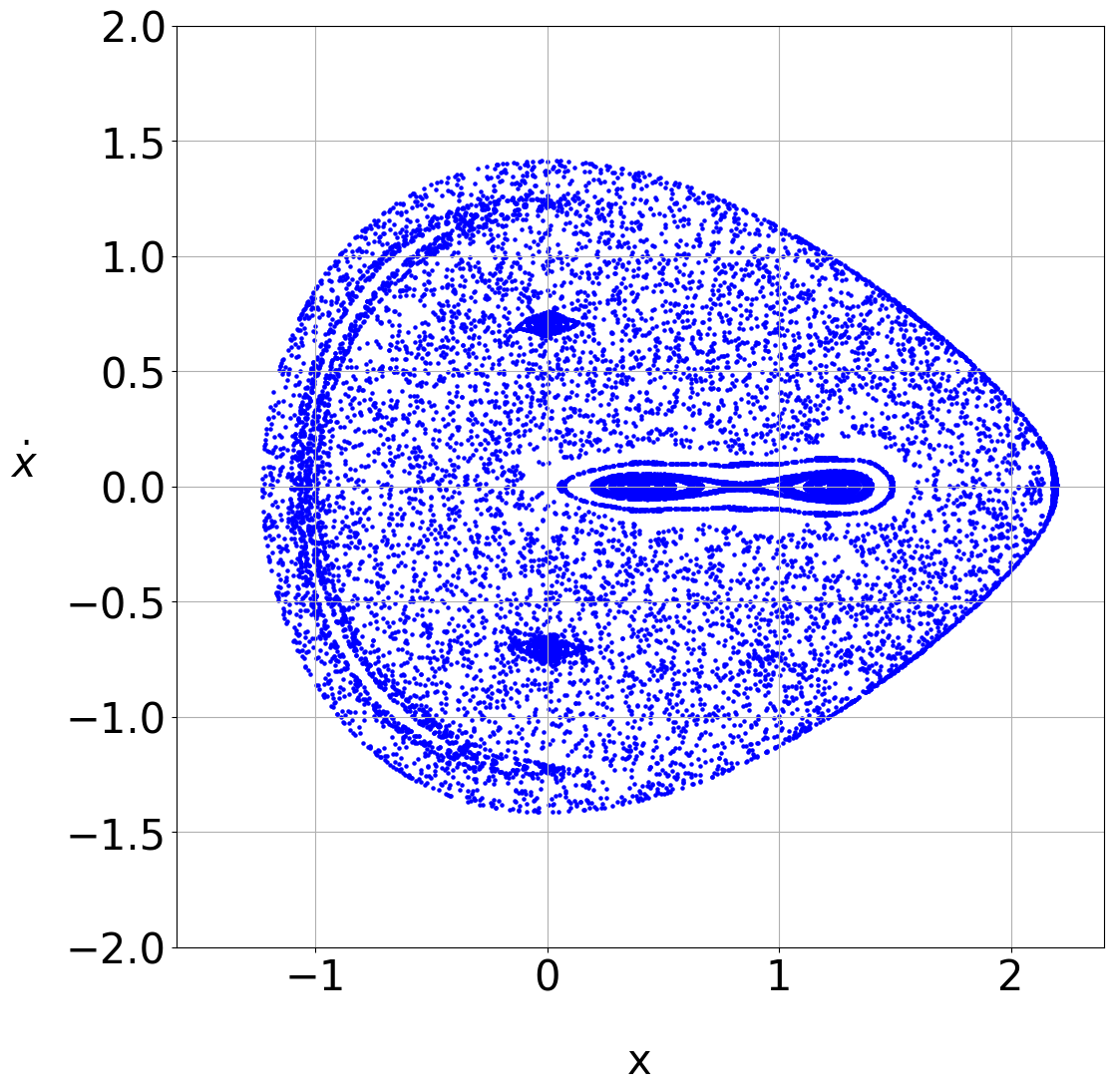}[d]\\
\includegraphics[scale=0.25]{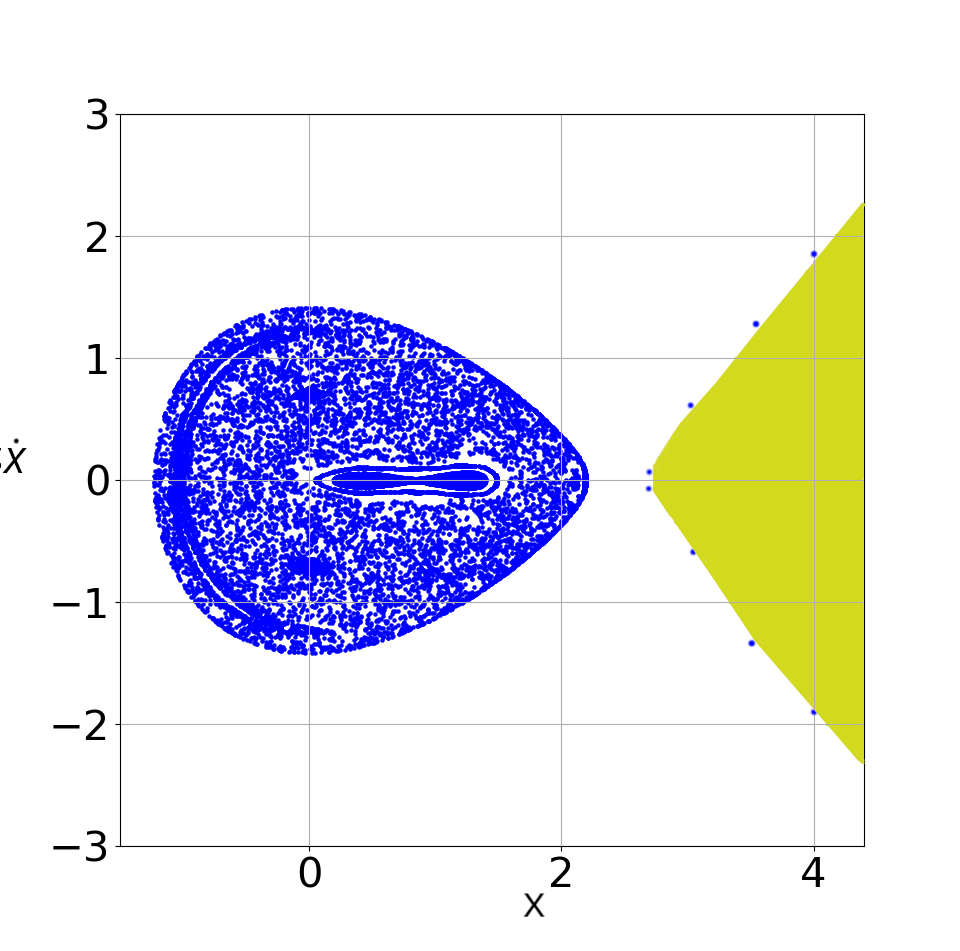}[e]
\includegraphics[scale=0.25]{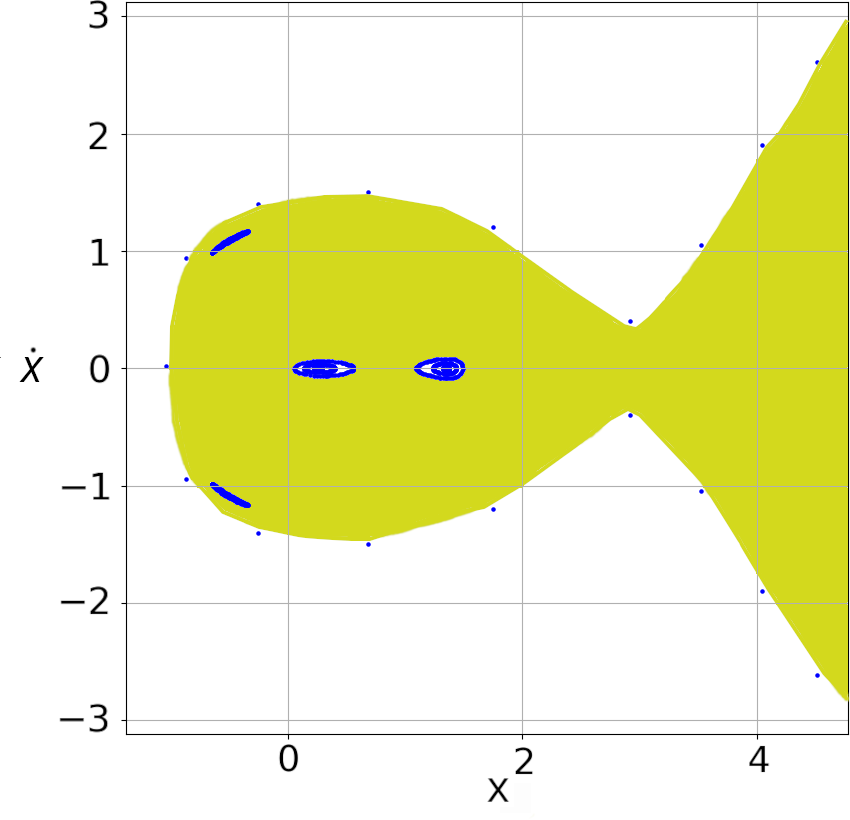}[f]
\includegraphics[scale=0.25]{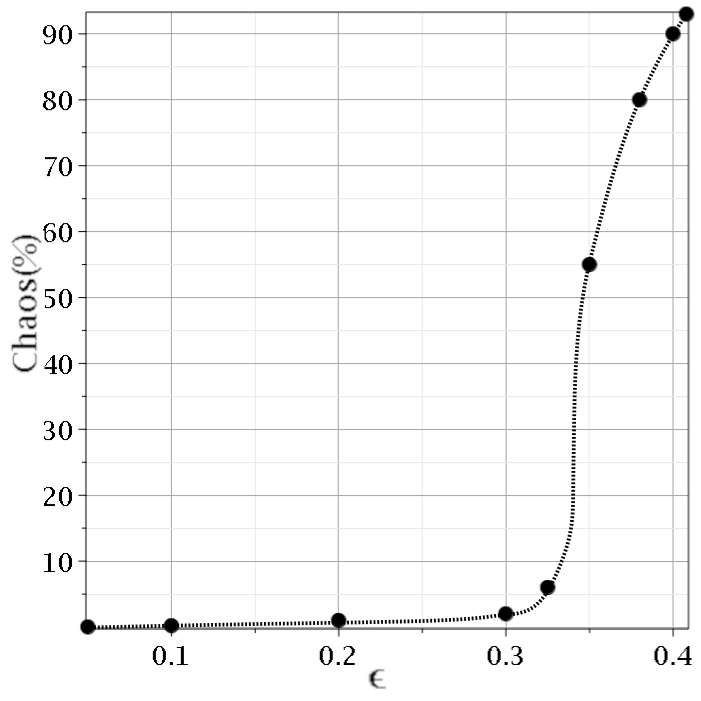}[g]
\caption{Poincaré surfaces of section $y=0, \dot{y}>0$ of the classical non integrable system  for a fixed value of the energy $E=1$ and for certain characteristic values of the coupling constant  $\epsilon$ that show the emergence of chaos and the introduction of escapes to infinity.  a) $\epsilon=0.05$ b) $\epsilon=0.3$ c) $\epsilon=0.35$. d) and e) refer to the same perturbation $\epsilon=0.4$  while f)  to $\epsilon=0.409$.The region of the escaping trajectories is coloured yellow. In g) we show the proportion of the chaotic orbits in the Poincar\'{e} surface of section as  a function of $\epsilon$. }\label{fig6}
\end{figure}

Hénon and Heiles \cite{henon1964applicability,henon1983numerical} gave Poincaré surfaces of section for $\epsilon=1$ and various values of the energy $E$. Here, however, we will fix $E=1$ (as in Section 2) and give surfaces of section ($y=0, \dot{y}>0$) for various values of $\epsilon$ that are characteristic for the onset of chaos as well as the introduction of escapes to infinity (Figs.~\ref{fig6}a,b,c,d,e,f). For small values of $\epsilon$ chaos is insignificant but beyond $\epsilon=0.3$ chaos increaes and becomes dominant (Figs.~\ref{fig6}bcd). Furthermore there is a second set of invariant curves on the right that extends to  +infinity. As $\epsilon$ increases this set approaches the central set of invariant curves and for $\epsilon_{crit}=0.4$ it joins it. The value $\epsilon=\epsilon_{crit}=0.408$ is the escape perturbation. For larger $\epsilon$ most trajectories escape to infinity. However there remain some small islands of stability (Fig.~\ref{fig6}f).

In Figs. 8abc  we see two main islands of stability around two stable points on the x-axis that belong to stable periodic trajectories. Between them there is an unstable periodic orbit and chaos as seen around it for $\epsilon\geq 0.3$ (Fig.~\ref{fig6}b). But in Figs.~\ref{fig6}cd we see many more islands of stability and between them unstable points and some small regions of chaos. The stable points on the left and on the right of Figs 6abc have become unstable in Fig.6d. In Fig. ~\ref{fig6}df several chaotic regions overlap and chaos becomes important. The case $\epsilon=0$ is the limit of Fig.~\ref{fig6}a when $\epsilon\to 0$. This is like Fig.~\ref{fig1}a.
rotated by 90 degrees. 

In Fig.~\ref{fig6}e we see that beyond the right limit of Fig.~\ref{fig6}d there is one more set of invariant curves that extend to $\infty$. As $\epsilon$ increases further to $\epsilon=0.409$ this set has joined the outermost invariant curve around the center and most chaotic trajectories escape to infinity on the right (Fig.~\ref{fig6}f).

%\subsection{Formal integral}

The proportion of chaotic trajectories as a function of $\epsilon$ is given in Fig.~\ref{fig6}g. For small $\epsilon$ this gives the chaotic area divided by the total area of the surface of section. We see that chaos is insignificant for $\epsilon$ up to $\epsilon=0.3$ but then it increases abruptly and reaches 92\% at the escape perturbation. Beyond the escape perturbation there remain some islands of stability that decrease gradually as $\epsilon$ increases. 

\begin{figure}[H]
\centering
\includegraphics[scale=0.222]{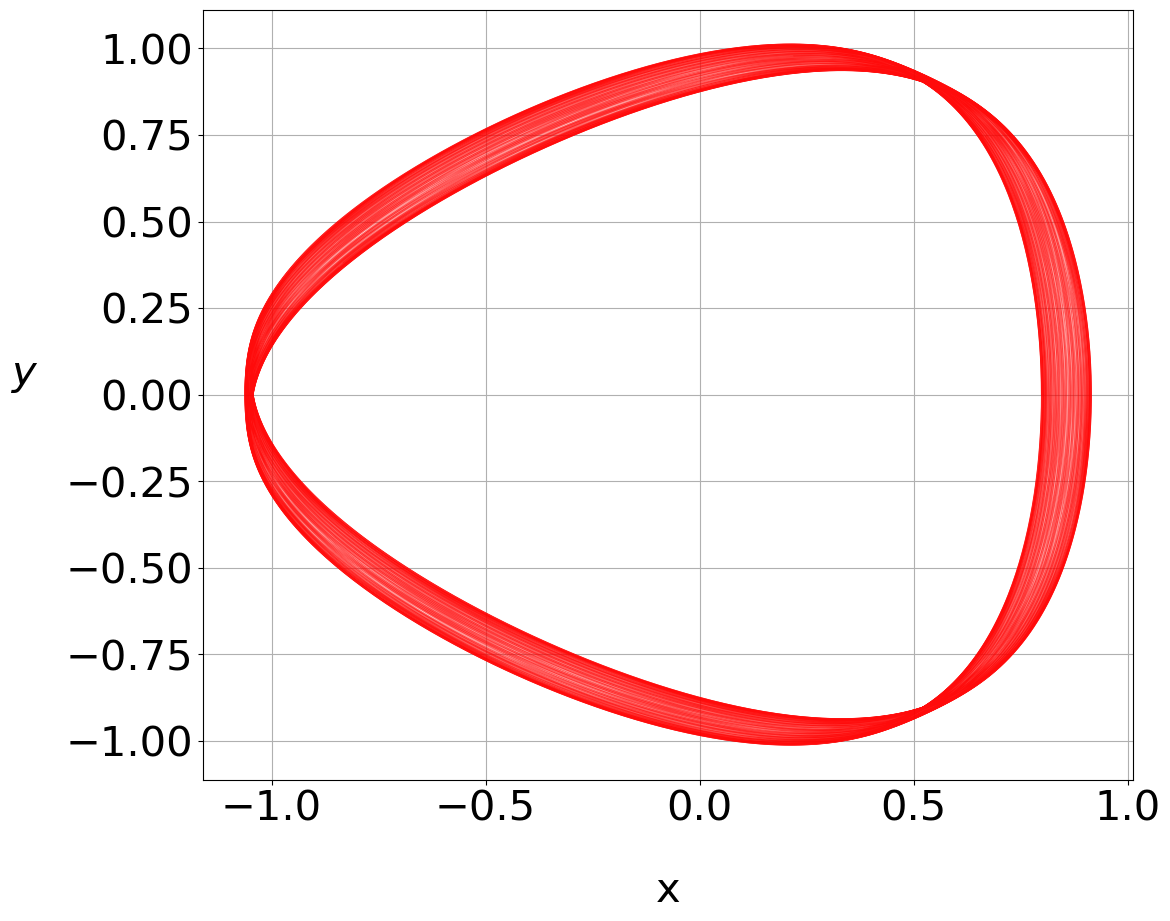}[a]
\includegraphics[scale=0.22]{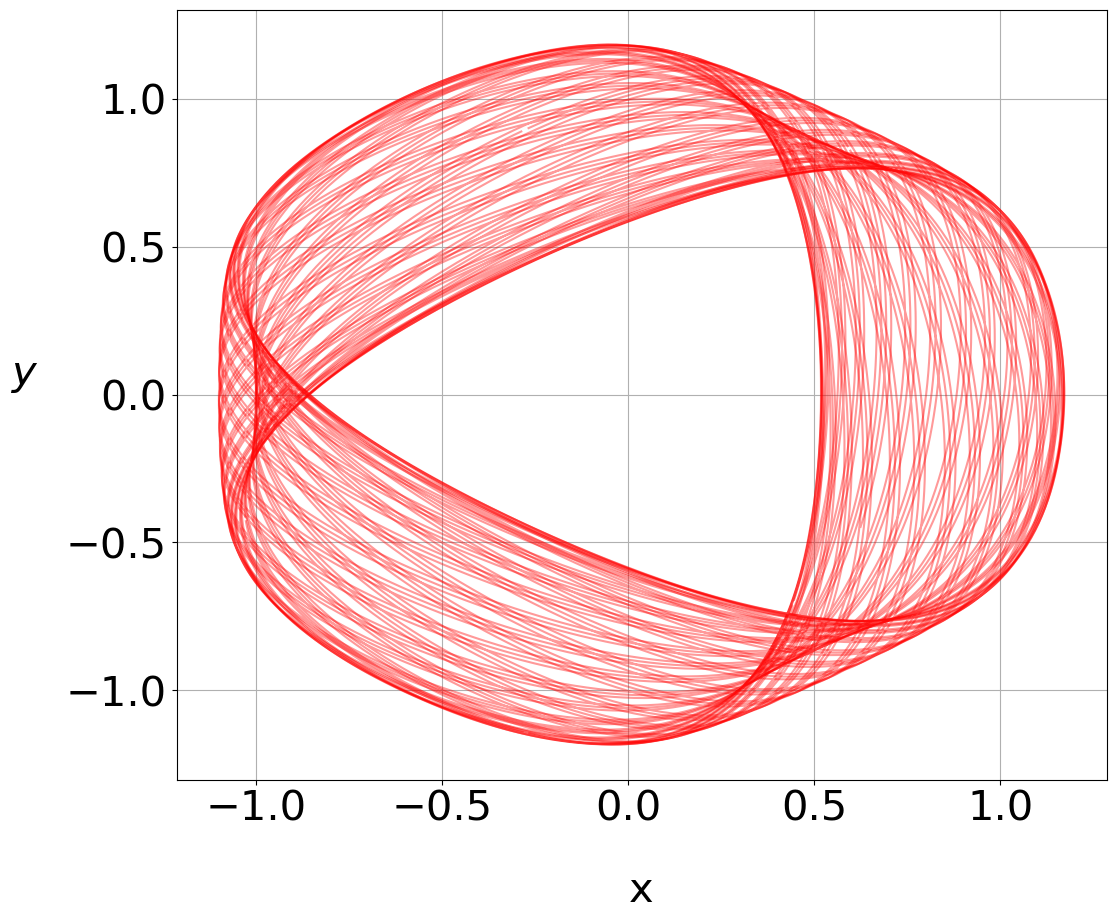}[b]
\includegraphics[scale=0.22]{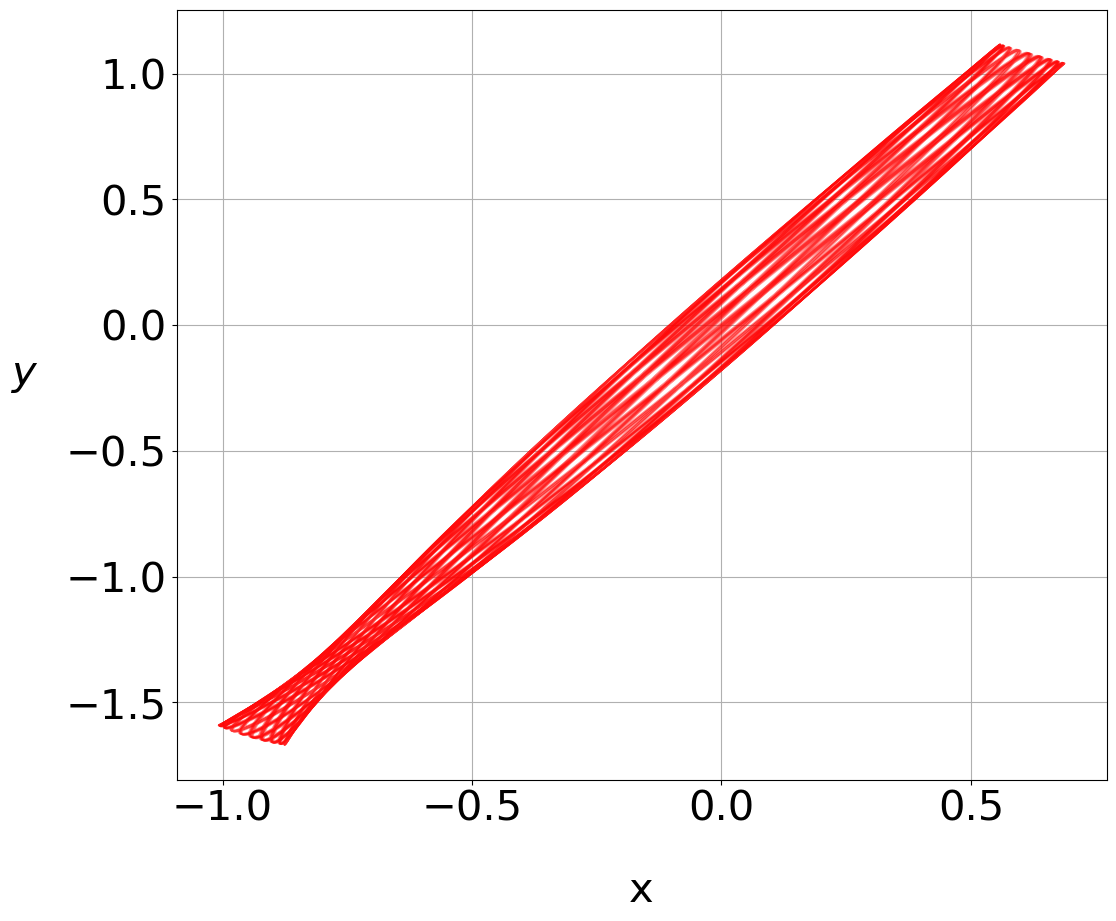}[c]
\includegraphics[scale=0.22]{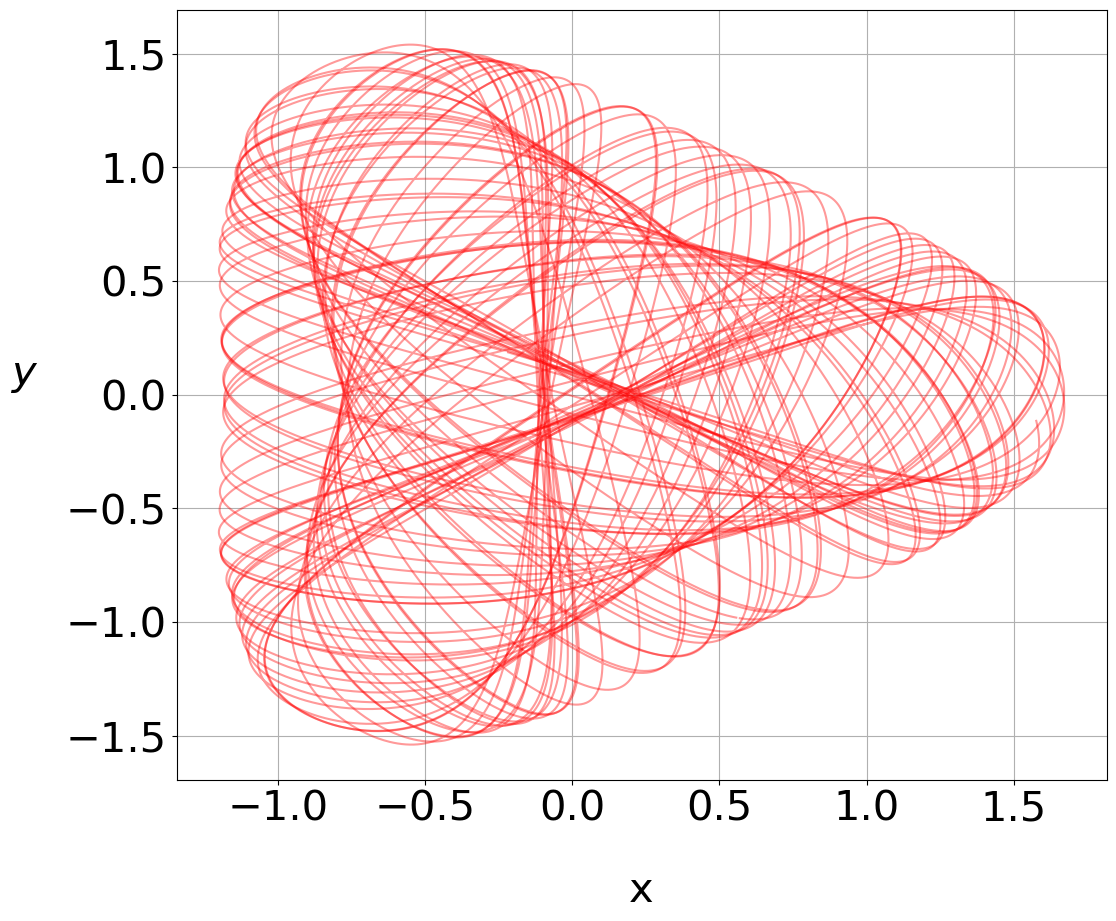}[d]
\caption{Four trajectories  around the Poincaré surface of section corresponding to $\epsilon=0.35$ and $E=1$: a) $x(0)=0.8, y(0)=0, \dot{x}(0)=0, \dot{y}(0)=1.2163$, b) $x(0)=-1, y(0)=0, \dot{x}(0)=0, \dot{y}(0)=0.8756$, c) $x(0)=0, y(0)=0, \dot{x}(0)=0.75, \dot{y}(0)=1.19896$ and d) $x(0)=-0.1, y(0)=0, \dot{x}(0)=0, \dot{y}(0)=1.41059$. The last one is chaotic.}\label{fig8}
\end{figure}

The forms of the classical trajectories are shown in Fig.~\ref{fig8}. There are trajectories forming rings around the stable periodic trajectories of Fig.~\ref{fig6} on the right and on the left of the origin (Figs~\ref{fig8}a,b), trajectories around the stable trajectory at $x=0$ and $\dot{x}>0$ (Fig.~\ref{fig8}c) and a similar trajectory for $\dot{x}<0$, and many chaotic trajectories as in Fig.~\ref{fig8}d. There are also small regions of trajectories around stable periodic trajectories of higher period.

These trajectories are very different from those of the integrable classical case (Fig.~\ref{fig1}e,f,g,h), except for the trajectories of Fig.~\ref{fig8}c above and below the center of Fig.~\ref{fig6}c, which are similar to the elongated Lissajous figure  of Fig.~ \ref{fig1}c.

\subsection{Quantum Case}

The Bohmian quantum analogue of the non integrable  Hénon Heiles system is very similar to the corresponding integrable case for the same value of the energy $E=1$. We work with the same wavefunction as in the integrable case. The corresponding average energies $E_{av}$ for increasing values of $\epsilon$ are shown in Fig.~\ref{fig12}a of the last section.

\begin{figure}[H]
\centering
\includegraphics[scale=0.22]{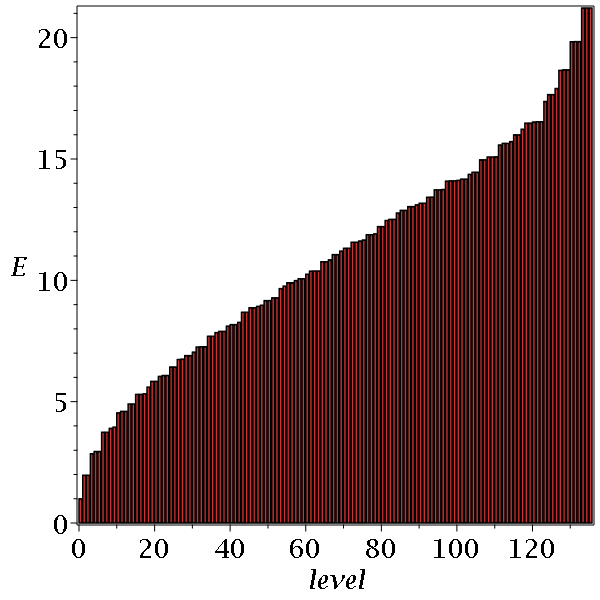}[a]
\includegraphics[scale=0.22]{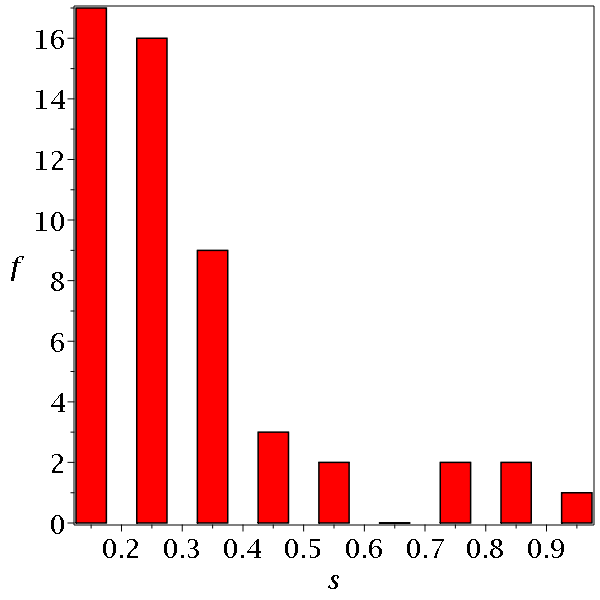}[b]
\caption{a) The distribution  of the first 136 energy  levels in the non integrable case when $\epsilon=0.2$. b) The histogram of the (135) successive energy gaps of value $s$.}\label{fig9}
\end{figure}

In Fig.~\ref{fig9}a we show the eigenvalues of this case, which are very similar to those of Figs.~\ref{fig3}a,b of the integrable case.  Figure~\ref{fig9}b gives the distribution of the differences of the adjacent eigenvalues, which is similar to Fig.~\ref{fig3}c of the integrable case.

\begin{figure}[H]
\centering
\includegraphics[scale=0.21]{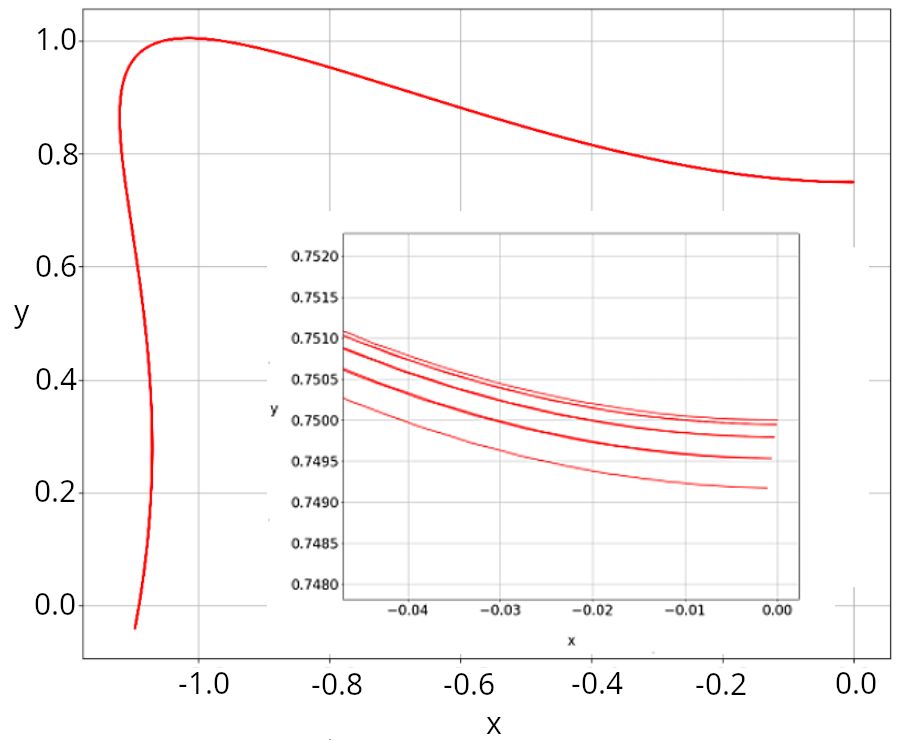}[a]
\includegraphics[scale=0.20]{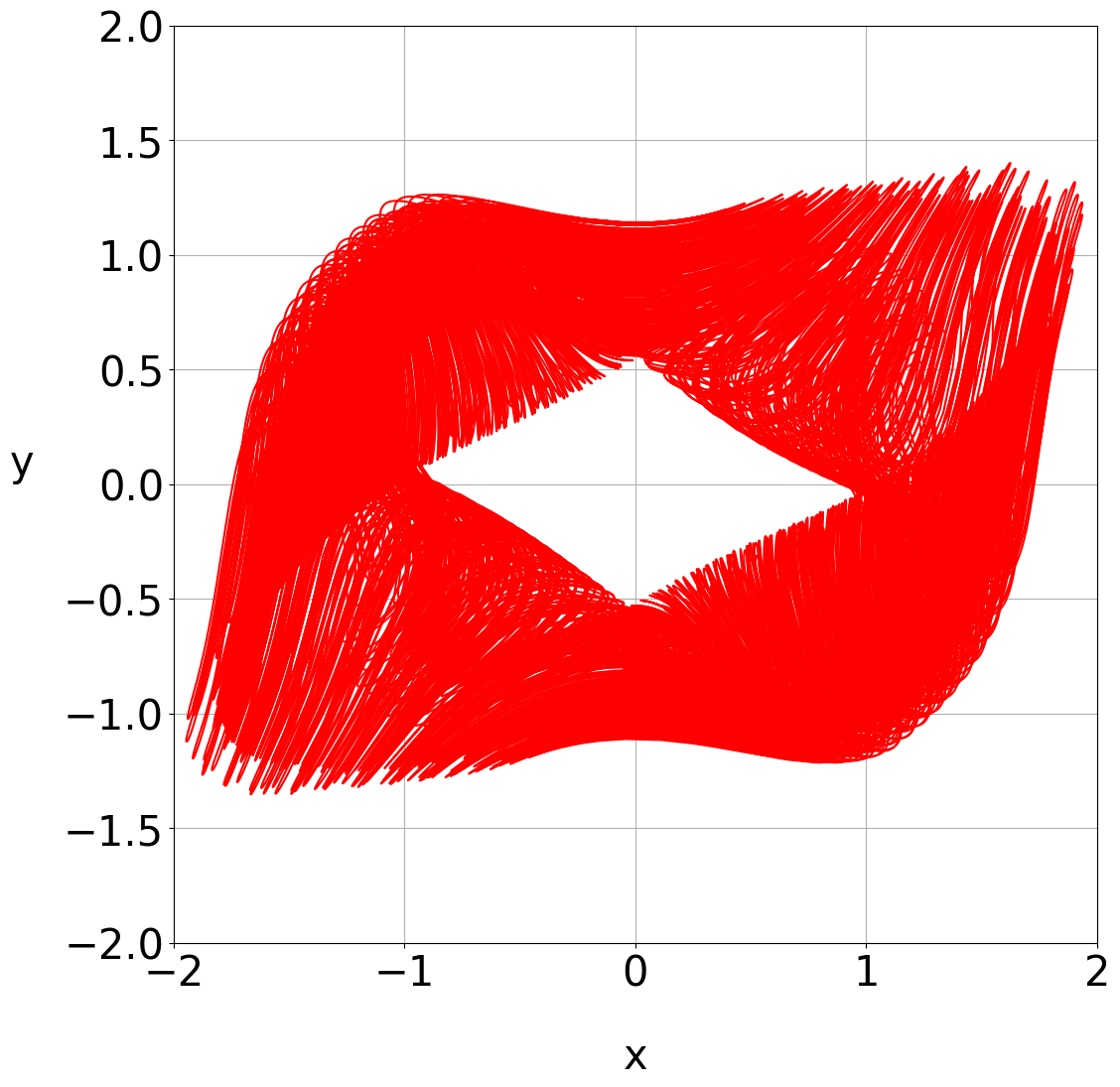}[b]
\includegraphics[scale=0.22]{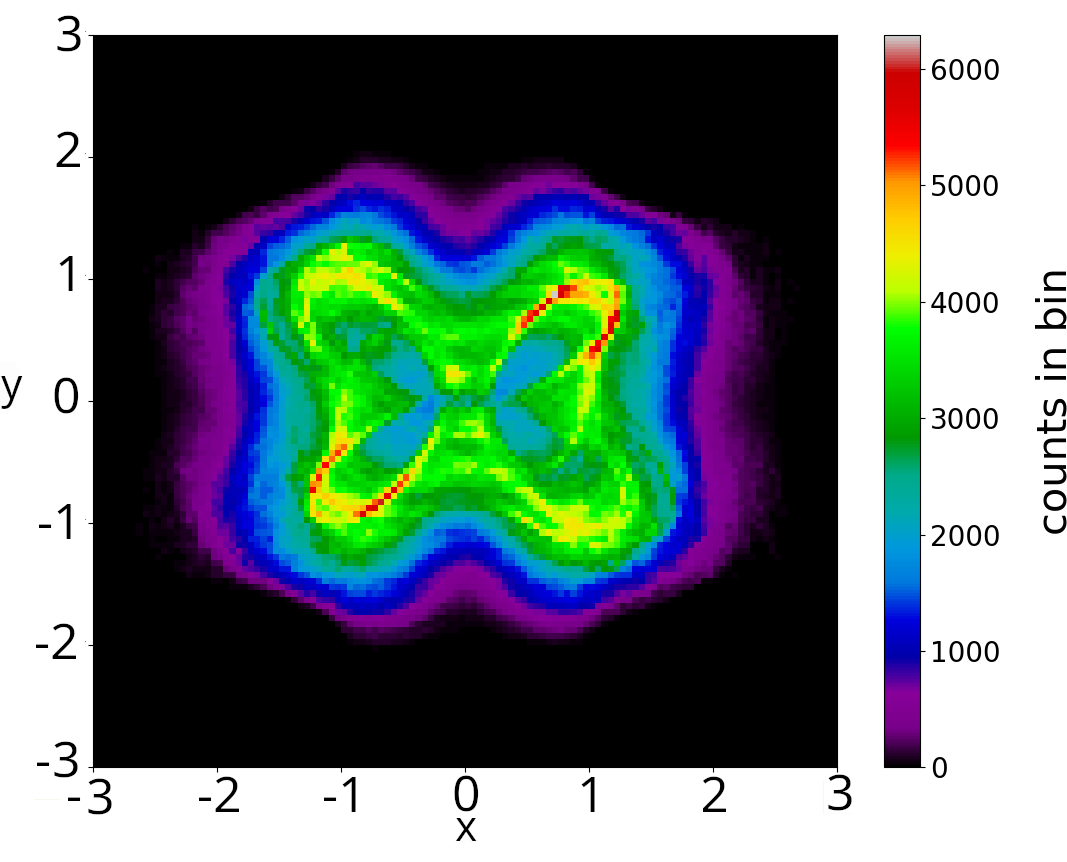}[c]
\includegraphics[scale=0.23]{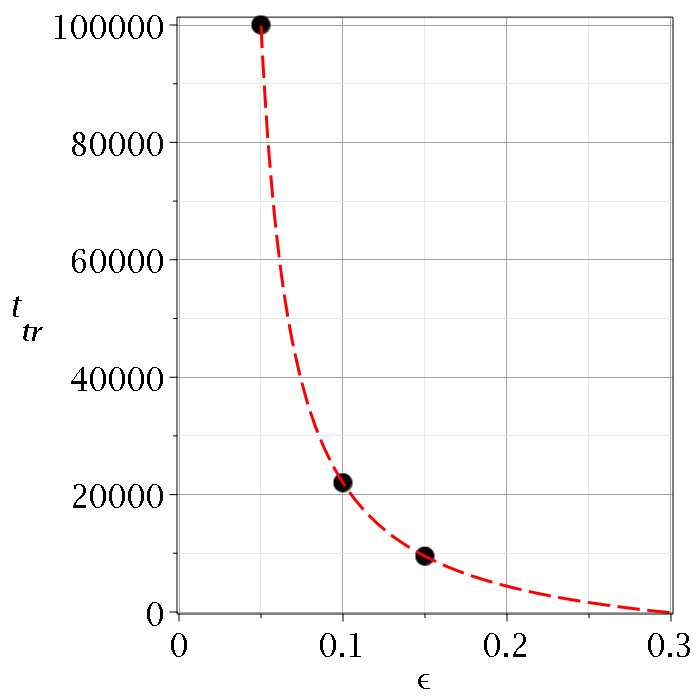}[d]
\caption{a) The Bohmian trajectory of the same initial condition of Fig.~\ref{fig4} for $\epsilon=0.05$:  a)  up to $t=4\pi$ (in the insert we show a zoom on a part of the trajectory of Fig.~\ref{fig10}a where we see that it is not periodic), b) up to time $t=10^3\times 2\pi$ and c) the colorplot of the long limit distribution of its points for $t=10^5\times 2\pi$. It is a chaotic-ergodic trajectory ($x(0)=0, y(0)=0.75$). d) The times needed for about $98\%$  of the trajectories, produced by  5000 Born-distributed initial conditions, to show their chaotic character for various values of $\epsilon$ (non integrable case).}\label{fig10}
\end{figure}

Fig.~\ref{fig10} shows a chaotic Bohmian trajectory  for $\epsilon=0.05$, and times a)  $t=4\pi$ b) $t=2000\pi$. Fig.~\ref{fig10}c is the colorplot of this trajectory for a long time ($t=10^5\times 2\pi$). These trajectories are similar to those of the integable case (Figs.~\ref{fig4}a,b,c). Finally,  Fig.~\ref{fig10}d gives the time needed for the trajectories of different $\epsilon$ to show their chaotic character and this is to be compared with Fig.~\ref{fig4}d. The chaotic character time is larger for smaller $\epsilon$ and tends to infinity as $\epsilon\to 0$. However, the times of the establishment of chaoticity are much larger in the non integrable case.  A detailed comparison between the integrable and non integrable case is given in the conclusions.

\section{Conclusions}

In this paper we compare the classically integrable and non integrable Hénon-Heiles systems both from a classical and a Bohmian quantum perspective. 

A) In the Bohmian approach the integrable and non integrable systems are similar in several ways.

\begin{figure}[!h]
\centering
\includegraphics[scale=0.24]{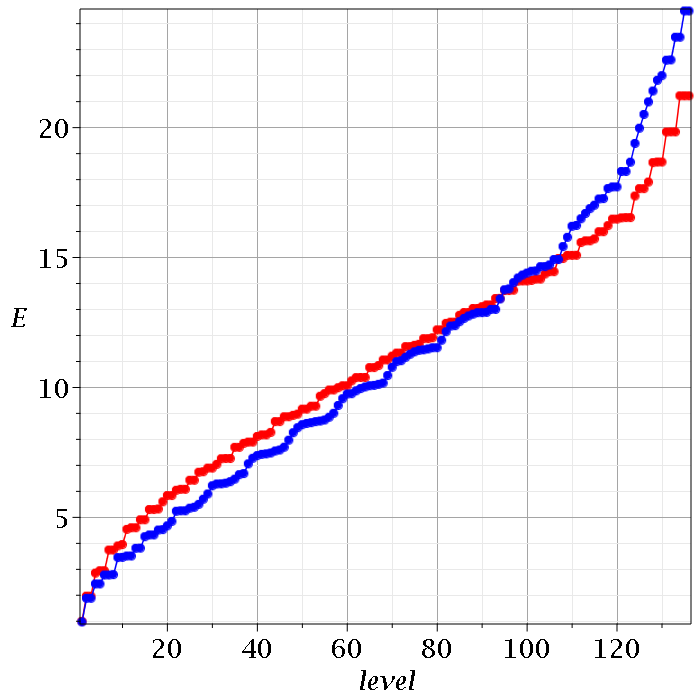}[a]
\includegraphics[scale=0.22]{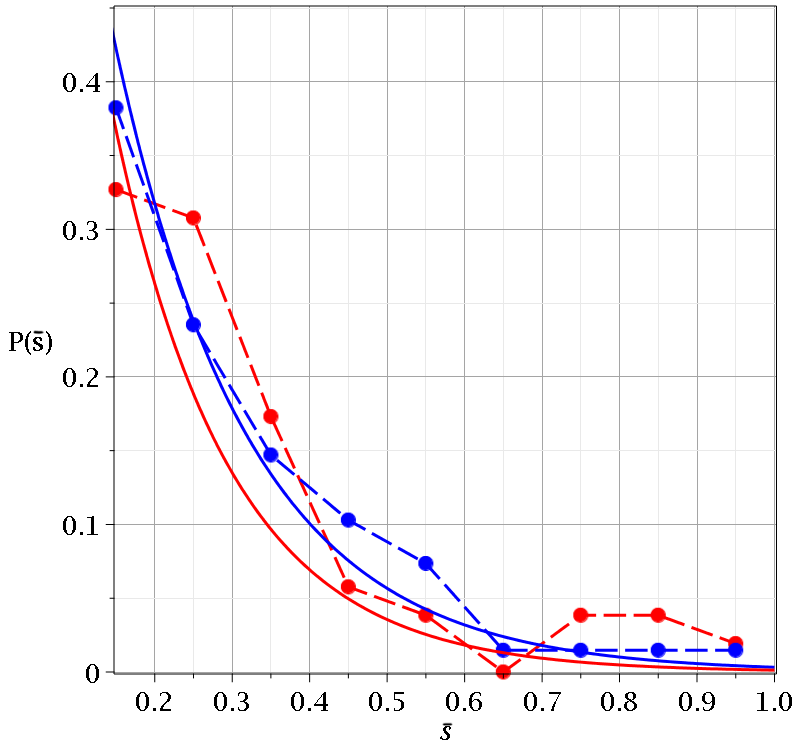}[b]
\includegraphics[scale=0.23]{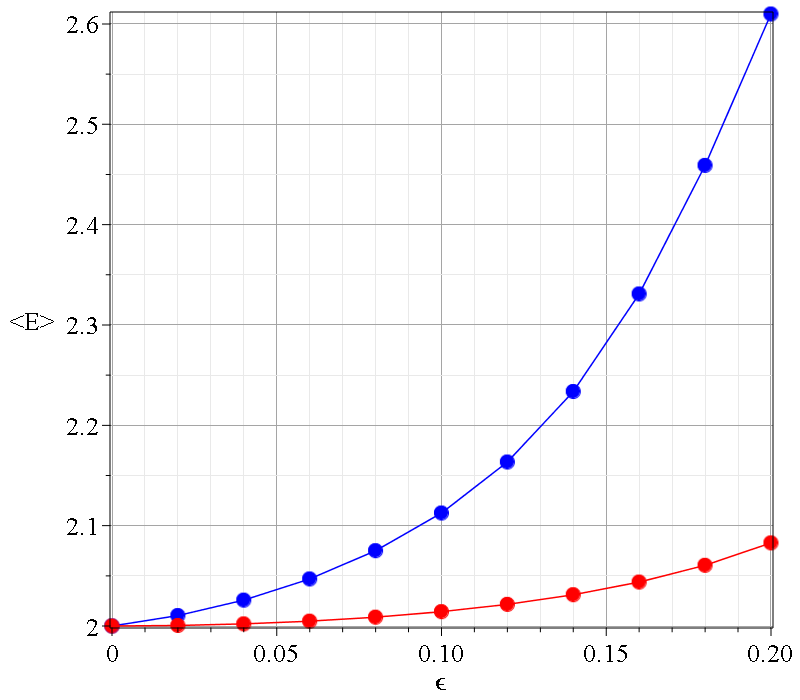}[c]
\includegraphics[scale=0.22]{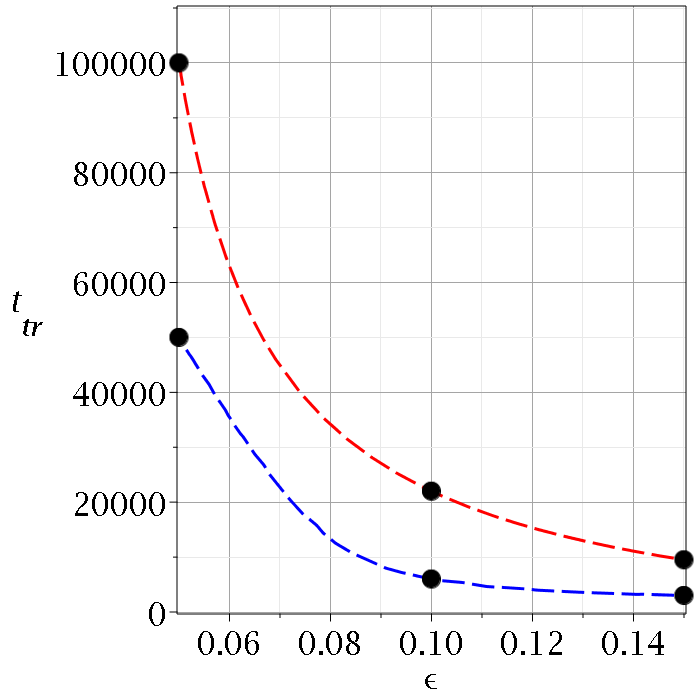}[d]

\caption{a) The distribution of the energy levels  of the integrable (blue) and the non integrable (red) case  and b) the probability of finding certain values for the spacings between successive energy levels  when  $\epsilon=0.2$. The solid lines are Poisson functions fitting the two observed distributions (dashed lines). We observe that the Poisson functions are very similar to each other. c) The average energy up to $t=10000$ as a function of $\epsilon$. The larger  $\epsilon$ leads to a larger  $\langle E\rangle$. d) The critical time when the majority of trajectories (about $98\%$) exhibits chaos as a function of $\epsilon$.  It is remarkable that in the integrable case (blue curve) Bohmian chaos emerges faster than in the non integrable case (red curve).}\label{fig12}
\end{figure}

\begin{enumerate}
\item The distributions of the eigenvalues of both integrable and non integrable systems are very similar (Fig.~\ref{fig12}a).
\item The energy differences $s$ between successive eigenvalues are similar. In Fig.~\ref{fig12}b we compare the distributions of $s$ in the integrable and non integrable cases (Figs.~\ref{fig3}c and \ref{fig9}c).
In both cases we have a Poisson distribution of the form
$P(\bar{s})\varpropto \exp(-\bar{s}),$
where $\bar{s}=s/s_{av}$ ($s_{av}$ is the average $s$) is the normalized gap  between successive energy levels.
\textbf{Namely, the Wigner function is not computable in the non integrable case of the H\'{e}non-Heiles system,} contrary to the common belief that distribution of the differences $\Delta E$ between adjacent eigenvalues has a Wigner form when the classical system is chaotic (while it has a Poissonian form in integrable cases). In fact, we find the same behaviour even for $\epsilon=0.35$ where the classical system has significant amount of chaos (Fig.~\ref{fig6}c). However, further work is needed in order to find whether this unexpected result is valid also in other cases with more complex Hamiltonians.
\item In Fig.~\ref{fig12}c we compare the average energies of the two systems for the same values ogf  $\epsilon$. The average energies start at $E_{av}=2$ for $\epsilon=0$ and increase as $\epsilon$ increases. However the increase is much slower in the non integrable case. The reason seems to be the following: the perturbation of the integrable case $\epsilon(xy^2+x^3/3)$ is much larger than the perturbation $\epsilon(xy^2-x^3/3)$ of the non integrable case for the same $\epsilon$.
\item The Bohmian trajectories are in general chaotic if $\epsilon\neq 0$ in both integrable and non integrable cases. The chaotic character of the trajectories takes more time to be established when the perturbation parameter gets smaller (Fig.~\ref{fig12}d). This time tends to infinity when $\epsilon\to 0$. However, the times for the same $\epsilon$ are larger in the nonintegrable case.
\item The colorplots (Figs.~\ref{fig4}c and \ref{fig10}c) are  similar. 
\end{enumerate}

	Therefore the Bohmian effects in both cases are similar but stronger in the integrable case.

B)  In the classical approach we calculate the invariant curves  for various values of $\epsilon$. 

\begin{enumerate}
\item In the integrable case the invariant curves form closed loops above and below the origin up around two stable points. But for $\epsilon$ above an escape value $\epsilon_{crit}$ the invariant curves starting close to the axis-x escape to the left, to minus infinity, surrounding also the closed invariant curves above or below the origin $(0,0)$. In the non integrable case there are closed invariant curves on the left and on the right of the origin, but there are also islands of stability symmetric with respect to the x-axis. If $\epsilon$ is larger than the escape value $\epsilon_{esc}$ most trajectories (but not all) escape to the right to infinity.

\item However, the main difference between the two systems is the appearance of chaos in the non integrable case. The proportion of chaotic trajectories is very small up to a critical value $\epsilon_{cr}$ and for larger $\epsilon$ it increases abruptly. Beyond the escape perturbation $\epsilon_{esc}$ there are only some small islands of stability that decrease gradually as $\epsilon$ increases further.

\item In the integrable cases the non escaping classical trajectories are in general Lissajous figures with axes parallel to the diagonals $y=\pm x$. In the limiting case $\epsilon=0$ the trajectories are closed loops or straight lines in limiting cases. The classical trajectories in the non integrable Hénon-Heiles caes are either ordered, surrounding in every case a stable periodic trajectory or chaotic. For example,  there are trajectories forming rings around the stable points with $y=0$ and $\dot{x}=0$. But if $\epsilon$ is larger than the escape perturbation $\epsilon_{crit}$ most classical trajectories escape to infinity.

\end{enumerate}

Our main conclusion is that  the Bohmian trajectories are chaotic, in general, both in the integrable and non integrable cases. These results are  important for the field of Bohmian chaos, which has been mainly developed by studying entangled wavefunctions of two non interacting quantum harmonic oscillators. There it was found that both ordered and chaotic trajectories coexist. However, as well as in the case $\epsilon xy^2$ that we studied in \cite{tzemos2024comparison}, we find that the interaction leads to the dominance of chaotic Bohmian trajectories in the course of time. Thus it is reasonable to expect that chaos is dominant in more general Hamiltonians.

The study of closed quantum systems with interacting parts is an important and necessary step before moving to the much more complicated, but also realistic, case of open quantum systems, where the enviromental interaction induces new effects as decoherence and dissipation \cite{gisin1992quantum,breuer2002theory,nassar2017bohmian}.

\section*{Acknowledgements}
This research was conducted in the framework of the program of the Research Committee of the 
 Academy of Athens “Study of order and chaos in quantum dynamical systems.” (No. 200/1026).

\bibliographystyle{elsarticle-num}

\bibliography{bibliography.bib}

\end{document}